\documentclass[3p,authoryear]{elsarticle}
\usepackage[utf8]{inputenc}
\usepackage{graphicx}
\usepackage{float}
\usepackage{hyperref}
\usepackage{amsmath}
\usepackage{subcaption}
\usepackage{xcolor}
\usepackage{soul}
\usepackage[all]{nowidow}
\usepackage{dcolumn}
\usepackage{xparse}

\usepackage[inline]{enumitem} 

\begin{document}
\title{Scale Effects in Ridesplitting: A Case Study of the City of Chicago}
\author[1]{Hao Liu\corref{cor1}}
\ead{hfl5376@psu.edu}
\author[2]{Saipraneeth Devunuri}
\ead{sd37@illinois.edu}
\author[2]{Lewis Lehe}
\ead{lehe@illinois.edu}
\author[1]{Vikash V. Gayah}
\ead{vvg104@psu.edu}
\cortext[cor1]{Corresponding author}
\address[1]{The Pennsylvania State University, University Park, PA, United States}
\address[2]{University of Illinois at Urbana-Champaign, Urbana, IL, United States}

\begin{abstract}
    Ridesplitting---a type of ride-hailing in which riders share vehicles with other riders---has become a common travel mode in some major cities. This type of shared ride option is currently provided by  transportation network companies (TNCs) such as Uber, Lyft, and Via and has attracted increasing numbers of users, particularly before the COVID-19 pandemic. Previous findings have suggested ridesplitting can lower travel costs and even lessen congestion by reducing the number of vehicles needed to move people. Recent studies have also posited that ridesplitting should experience positive feedback mechanisms in which the quality of the service would improve with the number of users. Specifically, these systems should benefit from economies of scale and increasing returns to scale. This paper demonstrates evidence of their existence using trip data reported by TNCs to the City of Chicago between January and September 2019. Specifically, it shows that increases in the number of riders requesting or authorizing shared trips during a given time period is associated with shorter trip detours, higher rates of riders being matched together, lower costs relative to non-shared trips, and higher willingness for riders to share trips.

\end{abstract}
\maketitle
\section{Introduction}\label{sec:intro}

\emph{Ridesplitting}, also called \emph{pooled service}, is a ride-hailing service in which customers agree to let the driver pick up and drop off other customers, who have placed separate and uncoordinated requests for the same service \citep{SAE2018,Shaheen2019a}. Launched in August 2014 \citep{Shaheen2018}, UberPool and Lyft Line (now Lyft Shared Rides) are the most common options in the US, while the TNC (transportation network company) Via offers \emph{only} ridesplitting in a handful of cities. Didi Chuxing's  ridesplitting option is called \emph{ExpressPool} \citep{Li2019}. Ridesplitting should be distinguished from app-based implementations of \emph{carpooling} or \emph{ridesharing}, such as Waze Carpool and Didi's Hitch\footnote{As \citet{Wang2019} explains, Hitch is targeted at drivers trying to defray the costs of commutes and intra-city trips.} option, insofar as the drivers---while not necessarily being professional or licensed chauffeurs---drive purely for pay; drivers do not share origins or destinations with any customers.

To carry out ridesplitting, a TNC matches customers and drivers in real time using routing and matching algorithms and travel time data. We will call trips \emph{shared rides} when they are carried out using a ridesplitting service. And we will call an ordinary ridehail trip a \emph{single ride}. We also distinguish \emph{authorized} shared trips, 
(also called \emph{requested shared trips}), (those which the user has given permission to share, by electing to place their request via UberPool, Lyft Shared Rides, etc.) from \emph{matched} shared trips (those which actually wind up being shared with another passenger who has placed a different request). A customer who authorizes sharing will not always wind up being matched.

With this vocabulary lesson finished, we arrive at the point of the paper. On matched shared trips, the customer almost certainly experiences a detour, because the vehicle must meander to pick up or drop off another customer. Hence, shared rides have longer expected travel times than single rides, and less travel time certainty (owing to the fact matches may be arranged after a journey is underway) \citep{Li2019}. The upside of sharing is that the TNC charges a lower fare. Some US cities and states (e.g., Chicago, New York City, Georgia, New Jersey) magnify the fare differential by charging lower excise taxes on shared rides than on single rides \citep{Lehe2021a}. Thus, ridesplitting offers a prime example of the classic tradeoff between money costs and travel time. \citet{Chen2017a} ranks travel time as the most important trip ``feature'' determining the choice of whether to ridesplit, and the next two features have to do with the trip's price.

In the US, the TNCs have offered ridesplitting only in large markets. Currently, \emph{Uber Pool} operates in just nine US cities and \emph{Lyft Line} is available in about fifteen cities. So apparently there is some threshold of potential demand necessary to operate ridesplitting in a financially sustainable way. This minimum scale is much larger than that necessary to maintain other, even fairly specialized, ridehail services---such as luxury black car service (offered in dozens of American markets). The argument of this paper is that ridesplitting, as an economic activity, exhibits two related characteristics:

\begin{itemize}
    \item First, \emph{economies of scale}: Economies of scale arise when the average cost of producing some unit of output falls with the level of output \citep{Silvestre1987}. Traditionally, economies of scale are associated with manufactured products such as steel, and  the focal cost is the producer's. For example, once a steel firm has built a blast or electric arc furnace, it is (per ton) cheaper to use it to make many tons of steel than a single ton. As regards passenger transportation, scholars have identified economies of scale on the \emph{user}'s side \citep{basso2010}. Consider a bus service, and let the units of output be bus trips, and figure into the ``costs of production'' the money value of the passenger's travel time (including time spent waiting at a stop). A famous result that follows from this way of looking at things is the so-called \emph{Mohring Effect}: \citet{Mohring1972} observes that when a higher ridership leads to higher service frequency (e.g., more buses per hour), the individual rider's wait times (at a stop) fall with total ridership.

          It is easy to see how ridesplitting might exhibit user-side economies of scale. If the density of requests for shared rides is very high, in both space and in time, then a TNC can match its customers together on trips with shorter detours. In the limit, as the density of requests for ridesplitting service in some market approaches infinity, anyone requesting a shared trip could be matched with a carload of others who have the exact same origin/destination pair; no detours would be needed.

    \item Second, \emph{increasing returns to scale}: ``A  technology exhibits increasing returns to scale if a proportionate increase in all inputs allows for a more than proportionate increase in outputs'' \citep{Vassilakis1987}. As with economies of scale, the canonical case is manufacturing, but increasing returns to scale is also observed in so-called ``search theories'' of the labor market \citep{diamond1982aggregate,Mortensen1994}. So conceived, the unit of output is a match, and the inputs to the matching process are the simultaneous stocks of employers searching for workers (i.e., posted job vacancies) and workers searching for employers (i.e., unemployed persons). Increasing returns to scale in matching arise whenever a proportionate increase to both stocks (i.e., a 10\% rise in both the number of vacancies and unemployed persons) begets a more-than-proportionate increase in the total flow of matches.

          It is not hard to imagine how increasing returns to scale in matching ought to also arise in ridesplitting. When the density of ridesplitting requests in a market rises, the only effect cannot be shorter detours on matched trips, because TNCs do not match 100\% of trips for which ridesplitting is authorized. If the detours required to match two trips would be too long, the passengers are simply taken to straight to their destinations as though they had authorized ``single ride'' trips. It follows that, when ridesplitting requests rise, the share of ridesplitting requests that are matched also rises.

\end{itemize}

In summary, we make several predictions about what ought to happen when the number of ridesplitting requests rises:

\begin{enumerate}
    \item the lengths of detours ought to decline
    \item the share of ridesplit requests that wind up matched ought to rise
\end{enumerate}

In addition, these effects might have two other potential knock-on effects. When the number of ridesplitting requests rises:
\begin{enumerate}
    \item The ratio between the fare of ridesplitting trips and the fare of ``single rides'' ought to decline (what \citet{Zhu2020} calls the ``fare ratio''), because the fare is based in part on the length of detours, and because the probability of a match rises.
    \item The proportion of all trip requests that authorize sharing ought to rise, because shared trips take less time, and because (potentially) their fares decline.
\end{enumerate}

This paper seeks to confirm that such behavior is observed in empirical data. The remainder of this paper is organized as follows. Section \ref{sec:data} describes the data that are used. Section \ref{sec:findings} presents the empirical findings. Finally, Section \ref{sec:conclusions} offers some concluding remarks.

\section{Data}\label{sec:data}

We use the ``Trips'' dataset reported by Uber, Lyft and Via (what Chicago calls ``Transportation Network Providers'') to the City of Chicago in 2019 to identify if economies of scale and increasing returns to scale are observed in real ridesplitting systems.\footnote{This dataset is publicly available at the following link: \url{https://data.cityofchicago.org/Transportation/Transportation-Network-Providers-Trips/m6dm-c72p}}  Each TNC provides the following information for each trip with an origin or destination within the City of Chicago:

\begin{itemize}[label=--]
    \item start and end timestamps associated with each trip, rounded to the nearest 15 min interval;
    \item the census tracts where trips began and ended;
    \item fare (\$), rounded to nearest multiple of \$2.50;
    \item additional charges such as taxes (\$);
    \item fees and other charges (\$);
    \item travel distance (mi);
    \item travel duration (min);
    \item whether sharing was ``authorized''; and,
    \item whether a shared trip was ``matched''.
\end{itemize}

Data cleaning consisted of removing certain records from the 2019 dataset. The process was guided by two goals.

\begin{enumerate}
    \item The first goal would apply to nearly any investigation of ride-hailing data: to remove potentially erroneous or trivial records. By \emph{trivial} we mean records which, while not necessarily recorded incorrectly, are not relevant for understanding anything about the nature of ride-hailing: for example, a trip someone cancels immediately upon entering the vehicle.
    \item The second goal is particular to what this paper's overall purpose: to compare variables at times which were similar, in order to minimize the impact of factors other than matching economies.  We describe each process in turn below.
\end{enumerate}

To remove erroneous or trivial records, we immediately filtered out trips with:
\begin{itemize}[label=---]
    \item travel times of less than 2 minutes;
    \item travel distances shorter than 0.1 miles;
    \item missing a pickup or dropoff census tract; and,
    \item trips that are not authorized as shared trips but coded as shared trips.
\end{itemize}

In addition, since we need to map the origins and destinations of all trips to the census tracts in the City of Chicago, trips with a pickup or dropoff census tract outside the City of Chicago were also excluded. Other studies in the literature, e.g., \citep{dean2021spatial}, also removed such trips in their analysis. However, \citet{dean2021spatial} focused on the factors contributing to sharing, and they used socioeconomic, built environment, temporal variables to develop two models that can be used to estimate the sharing decision in TNC trips. In addition, they only focused on the authorized shared trips whereas we use both authorized and matched shared trips.

In addition to the filtering process, systematic differences are detected between months before September and after October, so the 2019 dataset is divided two periods: January-September and October-December. The City of Chicago made a declaration on April 28, 2020\footnote{This declaration can be found at the following link: \url{http://dev.cityofchicago.org/open\%20data/data\%20portal/2020/04/28/tnp-trips-2019-additional.html}}: ``Due to data processing errors, some Transportation Network Provider trips in the first nine months of 2019 were not reported to the City of Chicago and therefore not initially included in the Transportation Network Providers - Trips dataset. The errors have now been corrected. We have received the trips and added them to the dataset.'' Although there is not a clear evidence to explain the continuous decrease in the number of shared trips observed in January-September, certain phenomena, such as extremely long detour distances being observed for short trips, in October-December are more suspicious and questionable. Unfortunately, there is not an assured theory to explain those patterns. Therefore, the data between October-December is excluded from this paper. A detailed analysis is provided in the Appendix \ref{sec:a1}. We do not consider trips from January 2020 onward due to an increase in ride-hailing taxes that occurred that month. The processed data to replicate the figures in the paper is available on Harvard Dataverse \citep{Liu2023}. The code is open-sourced and can be found at \url{https://github.com/UTEL-UIUC/Ridesharing-Scale-Effects}.

\begin{figure}[htb]
    \centering
    \includegraphics[width=4in]{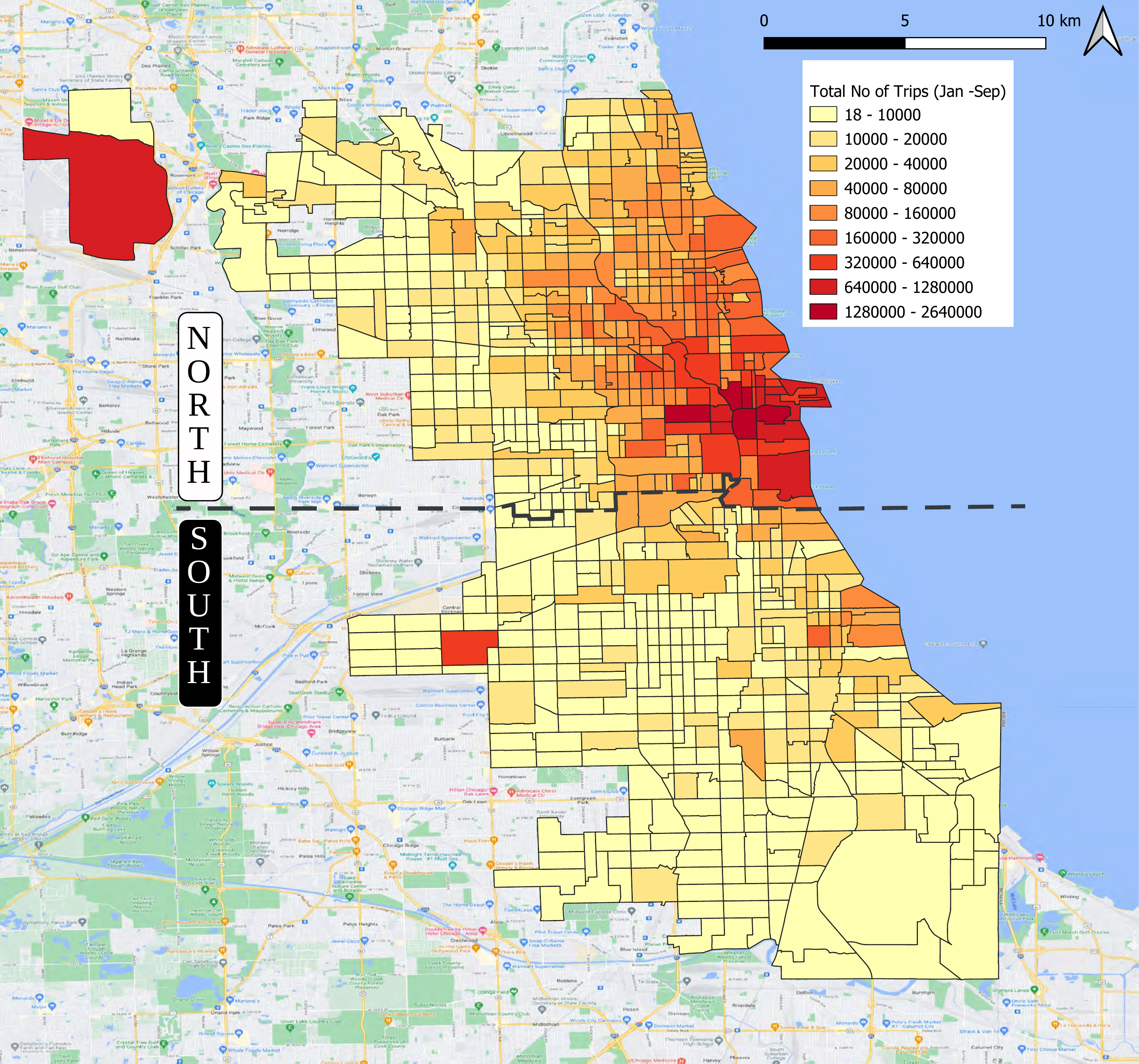}
    \caption{Distribution of trips, Jan-Sep 2019.}
    \label{fig:trip_map}
\end{figure}

The trip distribution in the dataset is shown in Figure \ref{fig:trip_map}. The heatmap represents the total number of trips starting from or ending at each census tract. Census tracts with very high number of trips either belong to Downtown Chicago (referred to hereafter as the Central Business District or CBD) or to the location of two major airports within Chicago (O'Hare and Midway). The concentration of trips drops as we move farther from the downtown unless the census tracts belong to special neighbourhoods such as touristic locations, universities, malls, etc. According to the trip distribution, we split the City of Chicago into two regions using the dashed line shown in Figure \ref{fig:trip_map}. We use this boundary to ensure the downtown area is located in the north side and try to make both regions have the similar number of cencus tracts. In the next section, WTS and matching rate in both regions are invesigated, separately.

We grouped the data based on hour ($h$), day of the week ($d$) and month ($m$), and the total/average number of (authorized shared/matched shared/single) trips in each hour-day-month combination was counted. For example, there are 8,813 matched shared trips between 9AM-10AM on all Mondays in March. Therefore, on average, there are 2,203 matched shared trips on a Monday between 9AM-10AM in March. Some metrics are defined based on this aggregated data, which will be explained when they are used. The unit fare per distance/time in a group is obtained by averaging the unit fares through all trips that belong to this group. Note a different aggregation level based on hour and day is used in Section \ref{sec:willingness} to explore the pattern of willingness-to-share (WTS) among weekdays. The definition of detour will be separately explained in Section \ref{sec:detour}. Table \ref{tab:para} provides a reference to the symbols used.

\begin{table}[htb]
    \setlength{\tabcolsep}{6pt}
    \caption{Notations}
    \centering
    \begin{tabular}{p{2cm}  p{8cm}  p{5cm}}
        \hline
        \hline
        \textbf{Notation}         & \textbf{Definition}                                                                                              & \textbf{Representation}                   \\
        \hline
        $n_{s_a,i}^{t(a)}(h,d,(m))$ & total(/average) number of authorized shared trips in hour $h$, weekday $d$ (and month $m$) in region i                     & Number of sharing authorized                         \\
        $n_{s_m}^{t(a)}(h,d,m)$   & total(/average) number of matched shared trips in hour $h$, weekday $d$ and month $m$                            & Number of sharing matched                        \\
        $n_{s,i}^{t(a)}(h,d,m)$    & total(/average) number of single trips in hour $h$, weekday $d$ and month $m$ in region i                                   & Number of single trips                        \\
        $n_{i}^{t(a)}(h,d,(m))$       & total(/average) number of trips (of both modes) in hour $h$, weekday $d$ (and month $m$) in region $i$                        & Number of total trips                              \\
        $\theta_{s_a, i}(h,d,m)$     & percentage of authorized shared trips in hour $h$ ,weekday $d$ and month $m$ in region $i$                                   & Willingness-to-share                    \\
        $\theta_{s_m, i}(h,d,m)$     & matched percentage of shared trips in hour $h$, weekday $d$ and month $m$ in region $i$                                                    & Sharing match rate                     \\
        $n^{O,D}_{s_m}(h,d,m)$        & average number of matched shared trips in pair (O,D) in hour $h$, weekday $d$ and month $m$                           & Number of sharing matched for each OD                     \\
        $d_{si(s_m)}^{O,D}(h,d,m)$        & average travel distance of single(/matched shared) trips in pair (O,D) in hour $h$, weekday $d$ and month $m$                           & Travel distance                           \\
        $de^{O,D}(h,d,m)$         & detour distance in pair (O,D) in hour $h$, weekday $d$ and month $m$                                             & Detour distance resulting from sharing trips \\
        $c_{si(s_m)}^{d(t)}(h,d,m)$       & average unit fare of single(/matched shared) trips in hour $h$, weekday $d$ and month $m$ based on distance (/time)                     & Unit fare                                 \\
        $e^{d(t)}(h,d,m)$         & unit fare ratio of shared trips to single trips in hour $h$, weekday $d$ and month $m$ based on distance (/time) & Cost-saving from sharing trips               \\
        \hline
        \hline
    \end{tabular}\label{tab:para}
\end{table}

\section{Findings}\label{sec:findings}
This section examines the presence of the economies of scale and increasing returns to scale, as introduced in Section \ref{sec:intro}, in ridesplitting. We show the TNC data supports the following conjectures: 

\begin{itemize}[label=---]
    \item WTS increases as the total number of TNC trips in the network rises;
    \item the detour lengths from sharing declines with the number of matched shared trips;
    \item  the match rate of shared trips increases with the proportion of authorized shared trips; and, 
    \item the unit fare ratio of shared trips to single trips decreases with the total number of TNC trips.
\end{itemize}

\subsection{Willingness-to-share}\label{sec:willingness}
WTS in region $i$ is defined as the fraction of trips authorized to be shared during a given hour-day combination in that region: 
\begin{equation}\label{eq:theta_sa}
    \theta_{s_a,i}(h,d) = \frac{n_{s_a,i}^t(h,d)}{n_i^{t}(h,d)}\times 100 \quad \text{(\% authorized shared trips)}.
\end{equation}

The WTS plotted as a function of the total TNC trips in both regions is shown in Figure \ref{fig:per_req}. The horizontal axes are the average number of trips of both modes in an hour across the same weekdays in the analysis period in region $i$, $n^{a}_i(h,d)$. The hour range in each plot is from 3AM to 2AM in the next day. The reason is that, as shown by Figure \ref{fig:per_req}, traffic volume reaches its minimum at 3AM in most weekdays in both regions. Therefore, if we consider the traffic evolution in a weekday as a cycle, we use this time point with minimal traffic volume as the start of a cycle. This is just for visualization and does not have impact on the increasing pattern, which is the main point of this section. 
\begin{figure}[H]
    \centering
    \begin{subfigure}{0.6\textwidth}
        \centering
        \includegraphics[width=\textwidth]{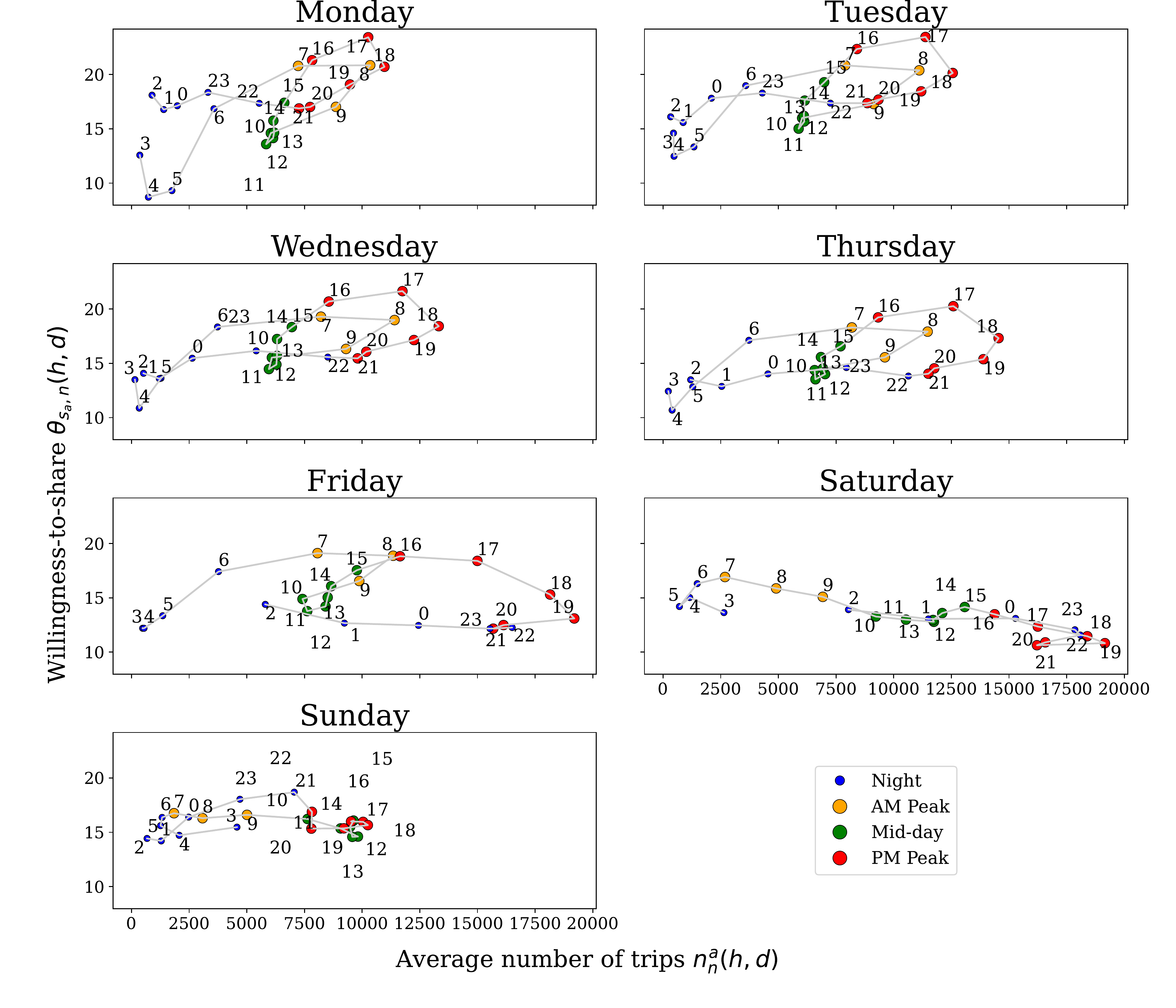}
        \caption{North side}
        \label{fig:per_req_north}
    \end{subfigure}
\end{figure}

\begin{figure}[H]\ContinuedFloat
    \centering
    \begin{subfigure}{0.6\textwidth}
        \centering
        \includegraphics[width=\textwidth]{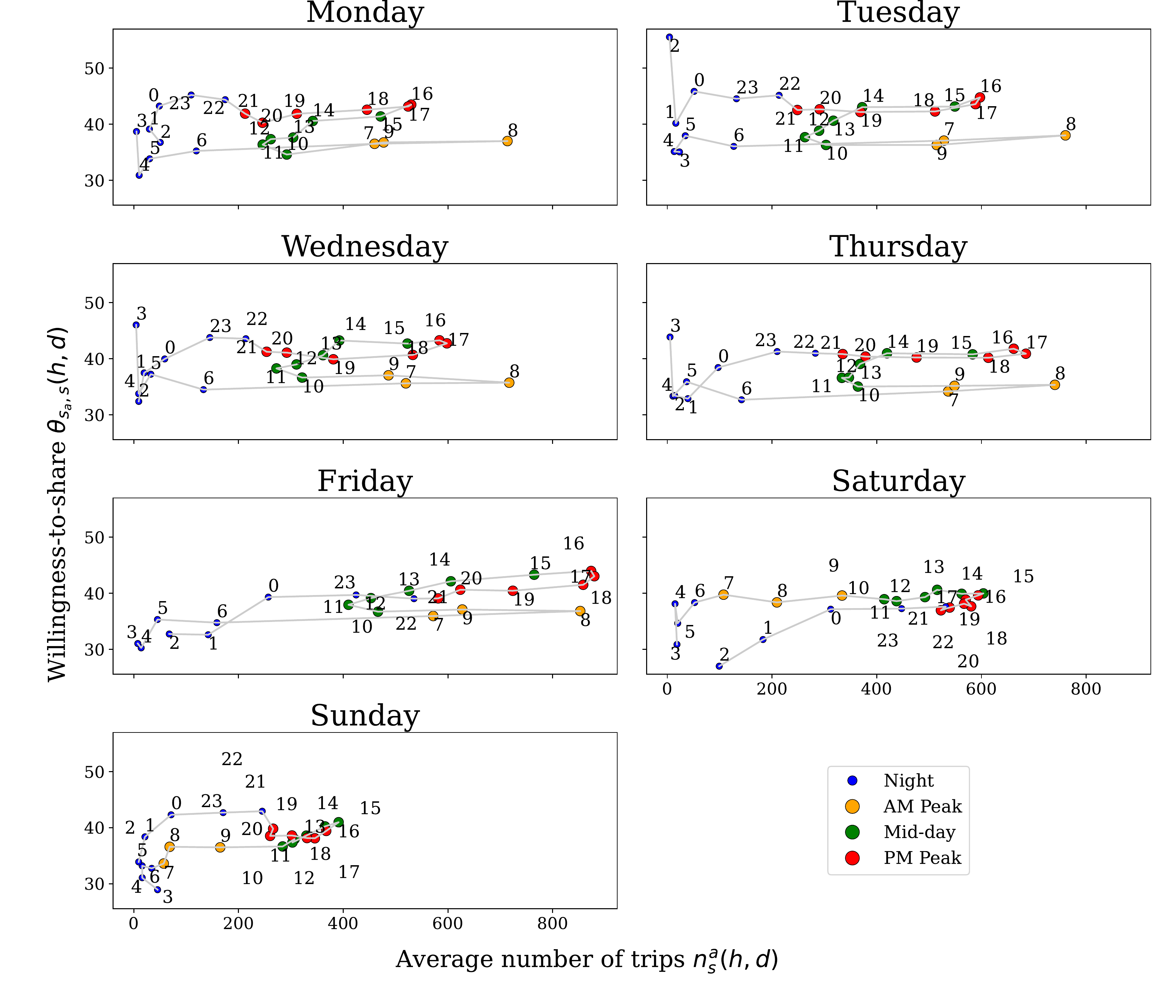}
        \caption{South side}
        \label{fig:per_req_south}
    \end{subfigure}
    \caption{Percentage of authorized shared trips.}
    \label{fig:per_req}
\end{figure}

In the north side region, WTS rises with the volume in the network on weekdays except for the night hours on Friday, while it is decreasing on Saturday and relatively stable on Sunday.

We then divided a day into 4 periods based on expected travel patterns on weekdays:
\begin{itemize}
    \item 7:00-9:00 --- AM Peak (in yellow)
    \item 9:00-16:00 --- Mid-day (in green)
    \item 16:00-21:00 --- PM Peak (in red)
    \item 21:00-7:00 --- Night (in blue)
\end{itemize}

The patterns in Monday through Thursday are similar in the north side region. The highest WTS appears between 16:00-18:00 while the WTS reaches its bottom in the early morning hours (4:00-5:00). Except for few outliers, such as 3:00-4:00 on Monday, WTS is increasing/decreasing with the rise/drop in the number of trips in other time. A  hysteresis pattern is also observed during the peak hours: the WTS during the start of peak hours is significantly higher than the value during the period in which the volume decreases. Friday presents a different pattern from other weekdays after the PM Peak Hours. Total trip frequency is higher and WTS is lower in those hours on Friday and weekends compared to the other weekdays. Specifically, the average WTS, which is defined as the number of total shared trips divided by the number of total trips, decreases from 17.3\% on weekdays to 13.6\% on weekends including after PM Peak hours on Friday in the north side region. The corresponding change, which decreases from 39.8\% to 38.8\%, is less significant in the south side region. The increased trip volume is likely due to an increase in social leisure trips during Friday PM evening hours. The lower WTS implies travelers are more likely to elect to ride alone to avoid detours during these activities. Another possible reason is that more people incline to share trips with friends and families using private vehicles on weekends as well as Friday nights, which leads to a higher average party size and a lower WTS for TNC trips.

Although the WTS in the north side region rises with the traffic volume on weekdays, the increasing trend is not observed in the south side region. On weekdays, the WTS in most hours fall into two stable levels: the early AM hours has a relatively low-level and stable WTS, while the PM hours and early night hours have a relatively high-level and stable WTS. Similar to the north side region, the pattern on weekends is distinct from weekdays. Moreover, although the south side region has a much lower traffic volume than the north side region, the average WTS is higher. We note, however, that the WTS is dependent on various factors, such as transportation system-related and demographic attributes. Therefore, a higher trip volume does not necessarily lead to a higher WTS if not controlling for such factors. However, we expect that  when the trip volume is high enough in a region, the WTS would increase with the trip volume if other attributes do not change. As Figure \mbox{\ref{fig:per_req}} shows, the average number of trips in the south side is less than 800 vph for most hours, which is considerably less than the value in the north side. A possible reason why it does not present an increasing trend is that the magnitude of the trip volume is too low to produce noticeable impact on the WTS.

\subsection{Detour}\label{sec:detour}
This section investigates the potential reduction of detour distances in shared trips resulting from the increase in the  number of shared trips occurring a given time. The detour distance between a given original-destination (OD) pair is defined as the average additional travel distance observed for shared trips during that time period compared to single trips between that OD within the same hour, which can be expressed as:

\begin{equation}\label{eq:detour}
    de^{O,D}(h,d,m)=d^{O,D}_{s_m}(h,d,m)-d^{O,D}_{si}(h,d,m)
\end{equation}

Therefore, trips between an OD pair within an hour are excluded from this analysis if only one mode is observed between that OD pair, since the detour distance could not be defined in these cases. There are 11,299,984 trips across 63,992 unique OD pairs that are retained after this filtering process. Of these, 8,680,382 are single trips and 2,619,602 are shared trips.

The relationship between the detour distance and the number of matched shared trips in the 16 OD pairs with the highest number of trips is examined. This includes 1,261,302 single trips and 118,733 shared trips in total. For brevity, only the results for the first 6 OD pairs are shown in Figure \ref{fig:detour_dis_3}, and the results for the other 10 OD pairs are shown in Appendix \ref{sec:a2}. Since the number of data points for each OD pair is large and most points have a relatively small number of shared trips, the cloud of points in classical scatter plots is very dense at the region with small values of shared trips, which makes obtaining visual insights from the plots impossible. Instead, binned scatter plots are used, which  can provide visual assessment of features such as monotonicity more clearly. Figure \ref{fig:detour_dis_3} shows the data using binned scatter plots created using the R package "binsreg" \citep{cattaneo2019binscatter, cattaneo2019binscatter2}. For each OD pair, we use quantile binning with 8 bins, which tries to assign the same number of observations into 8 bins. The point and bar in each bin correspond to the point estimates evaluated at the mean and the 95\% confidence interval(p-value 0.05) for the detour distances observed within each bin. The shaded areas provide a confidence band for both variables associated with each bin. Figure \ref{fig:detour_dis_3} shows the detour distance generally decreases with the number of shared trips for all OD pairs. The trend is more obvious when the number of shared trips is low. On the contrary,  the decreasing trend is not as clear when the number of shared trips is very large.

\begin{figure}[H]
    \centering
    \begin{subfigure}{0.45\textwidth}
        \centering
        \includegraphics[width=\textwidth]{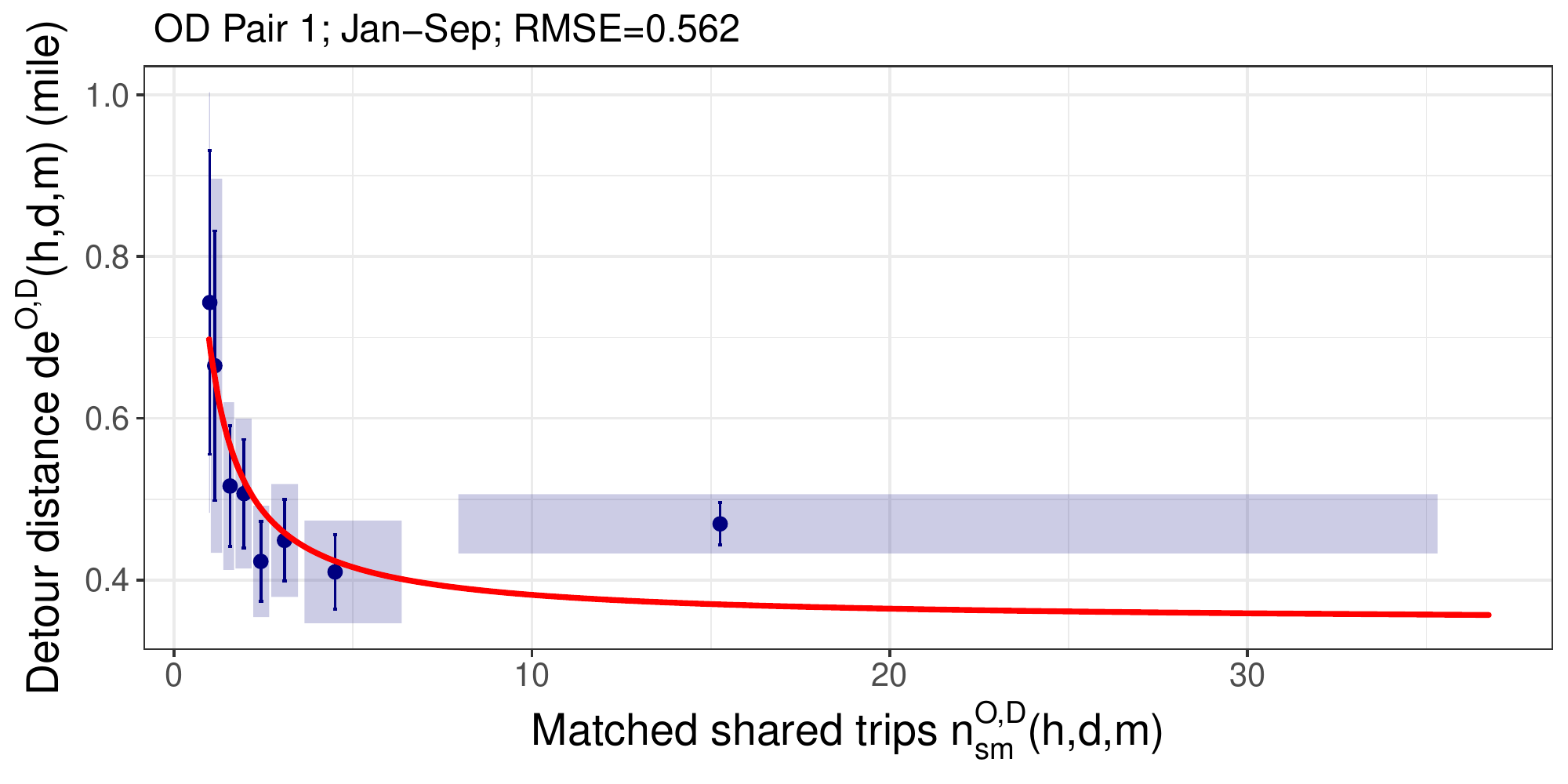}
    \end{subfigure}
    \begin{subfigure}{0.45\textwidth}
        \centering
        \includegraphics[width=\textwidth]{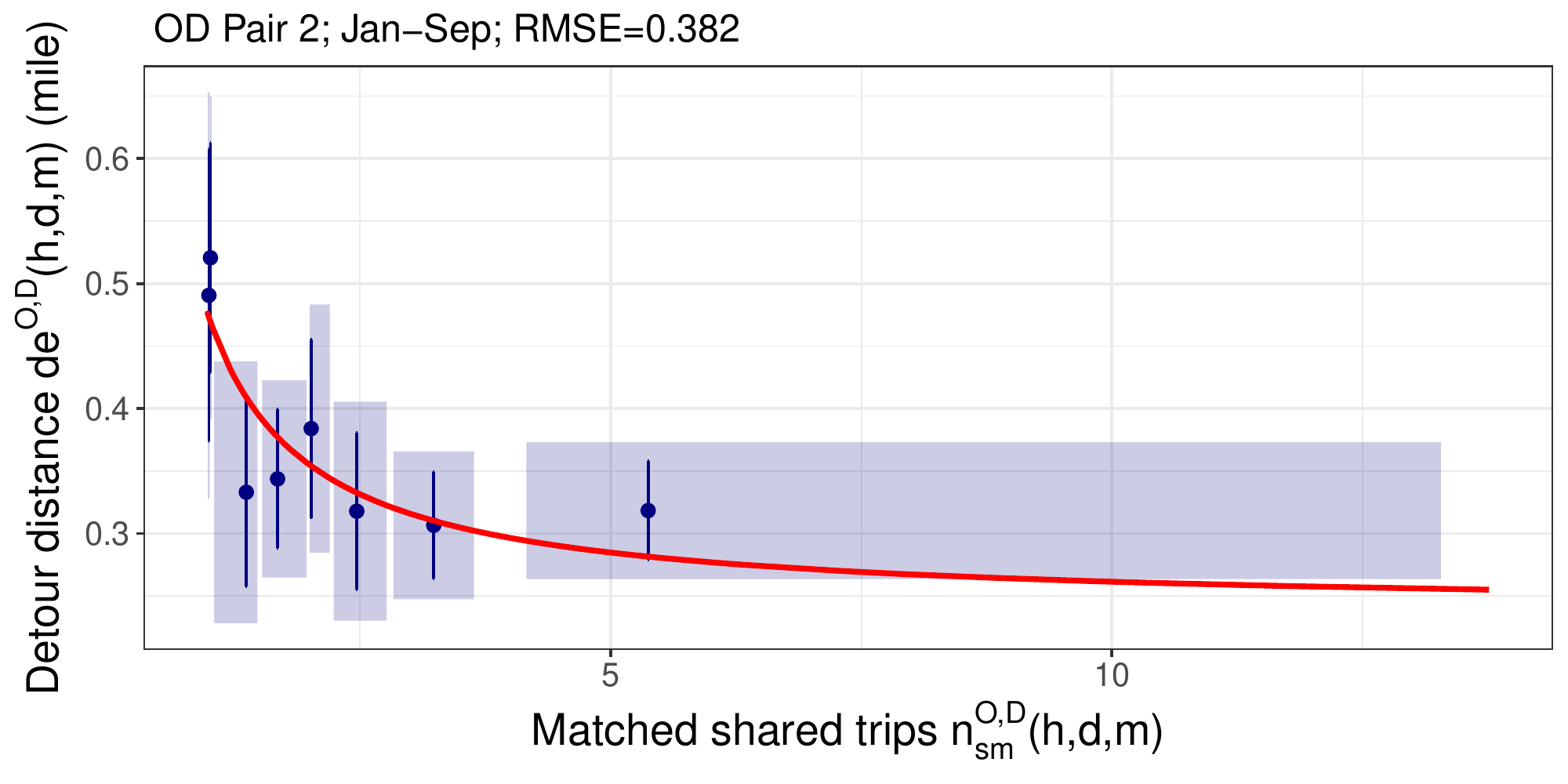}
    \end{subfigure}

    \begin{subfigure}{0.45\textwidth}
        \centering
        \includegraphics[width=\textwidth]{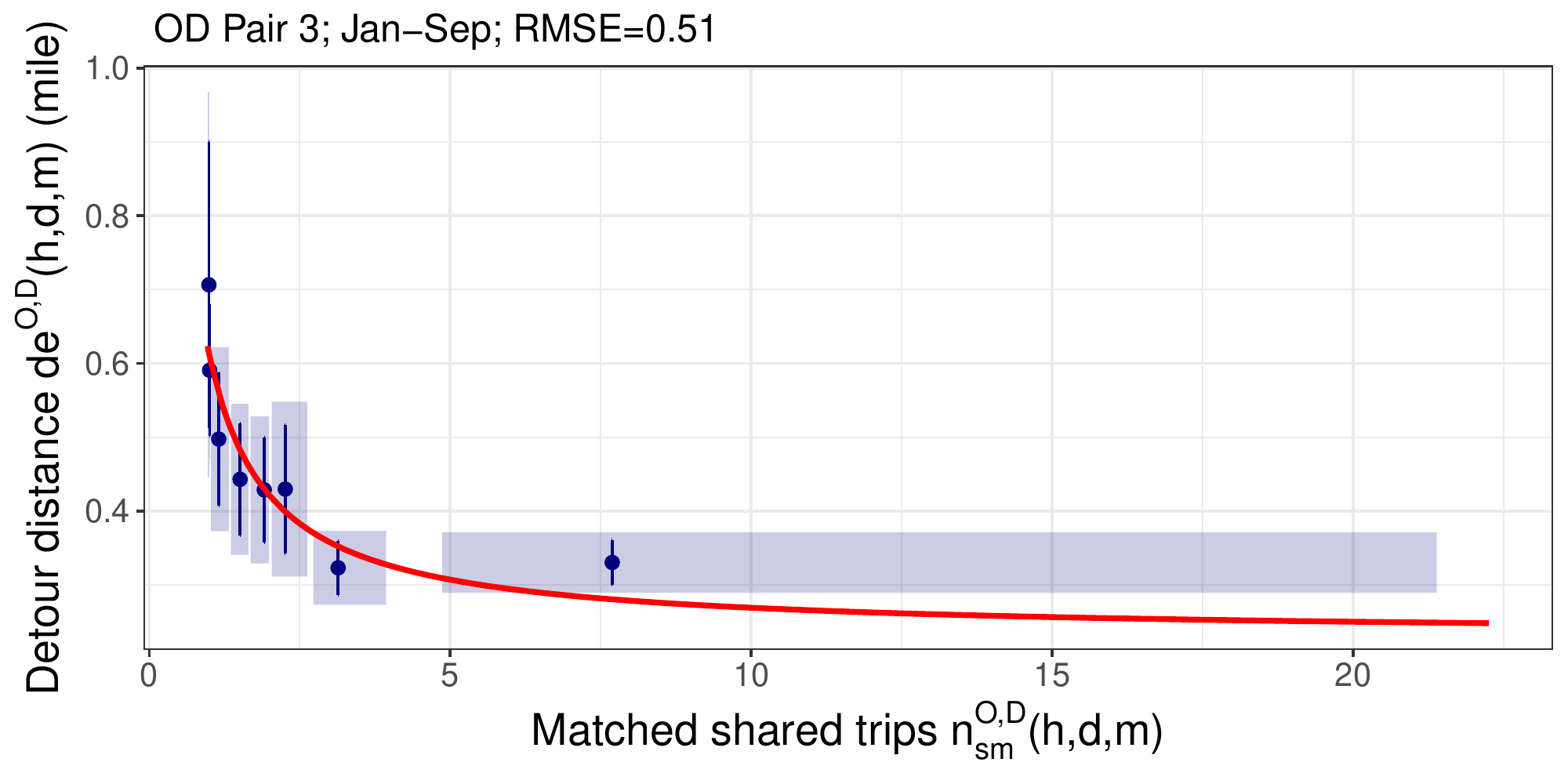}
    \end{subfigure}
    \begin{subfigure}{0.45\textwidth}
        \centering
        \includegraphics[width=\textwidth]{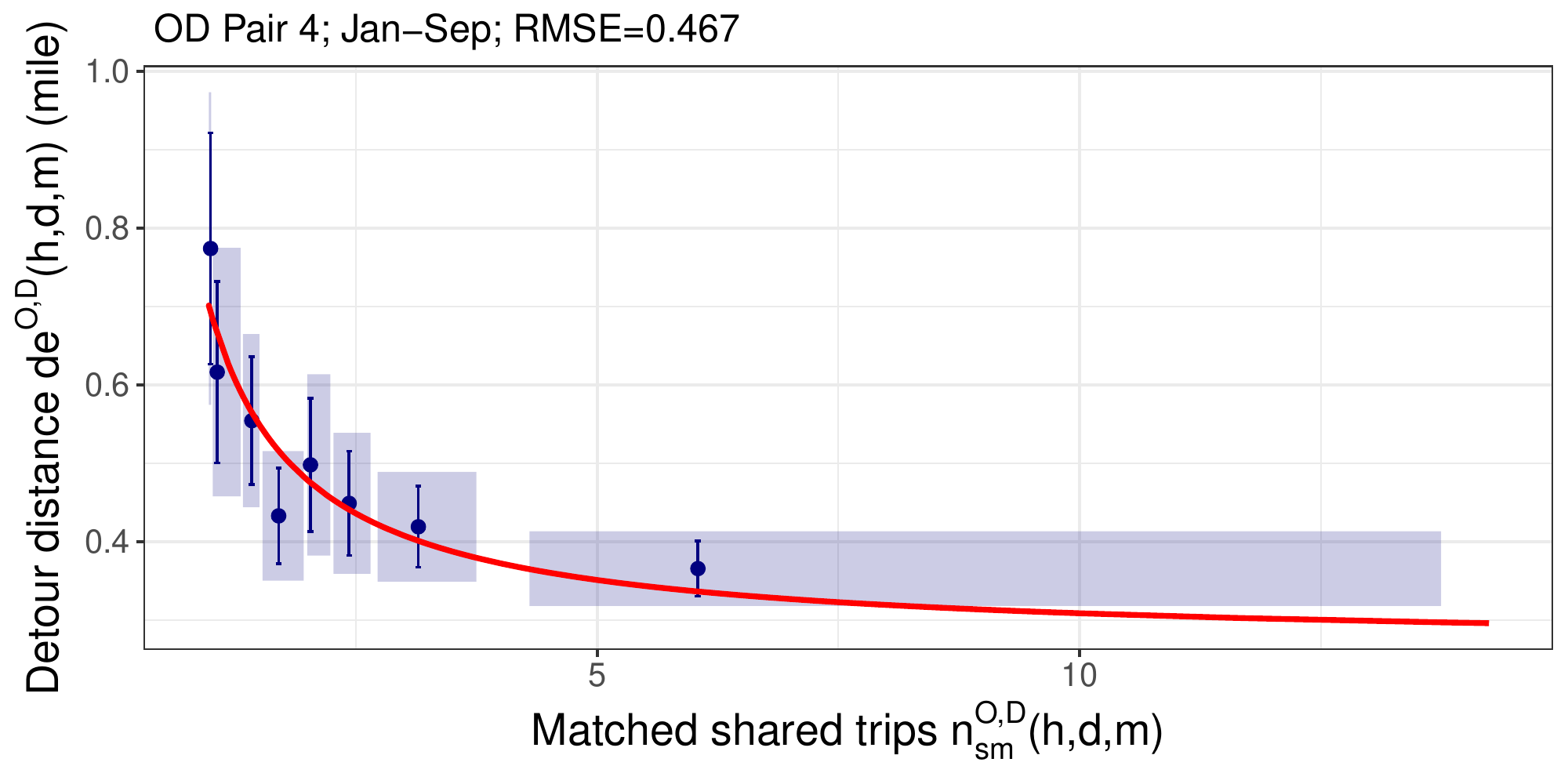}
    \end{subfigure}

    \begin{subfigure}{0.45\textwidth}
        \centering
        \includegraphics[width=\textwidth]{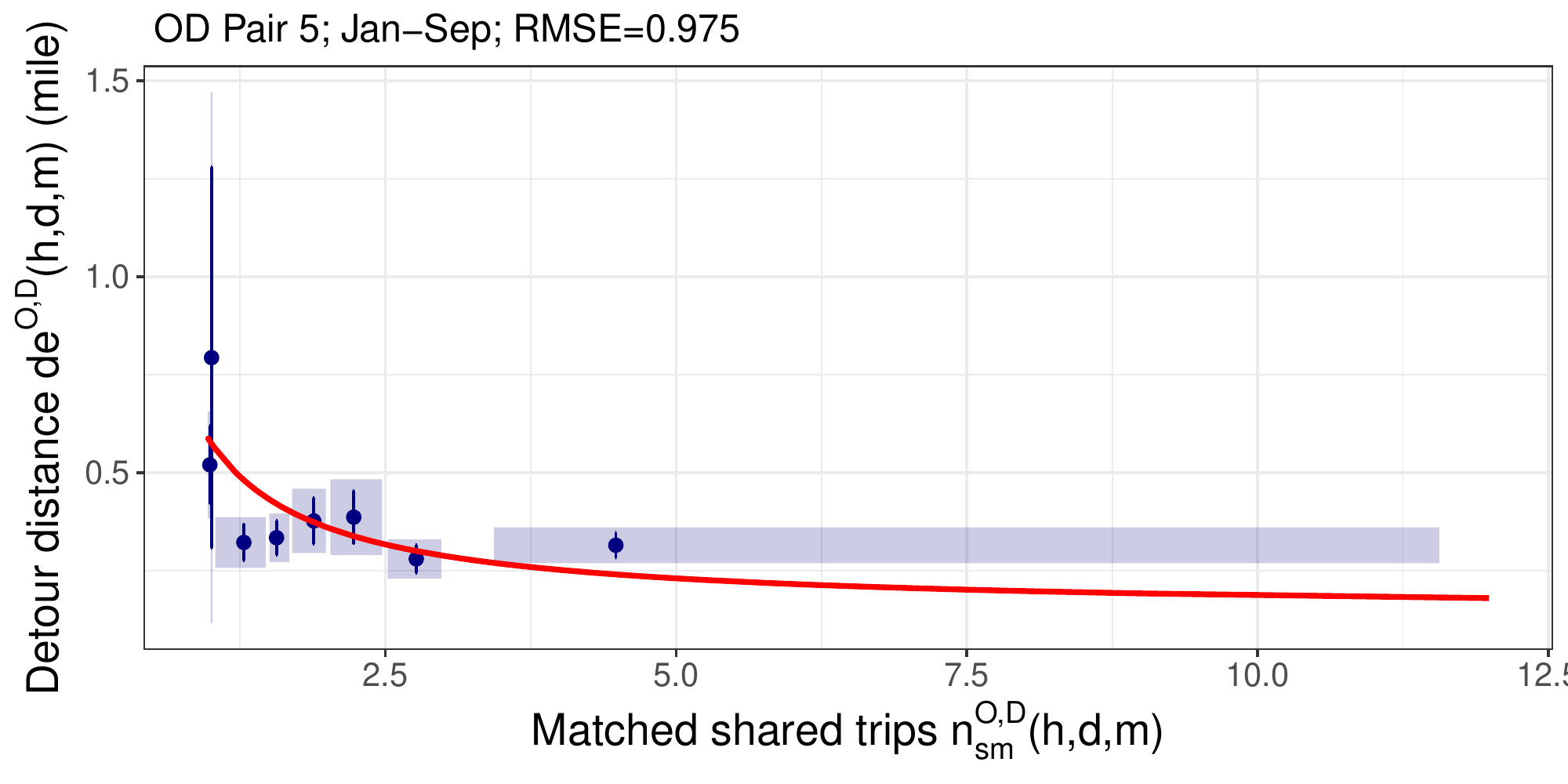}
    \end{subfigure}
    \begin{subfigure}{0.45\textwidth}
        \centering
        \includegraphics[width=\textwidth]{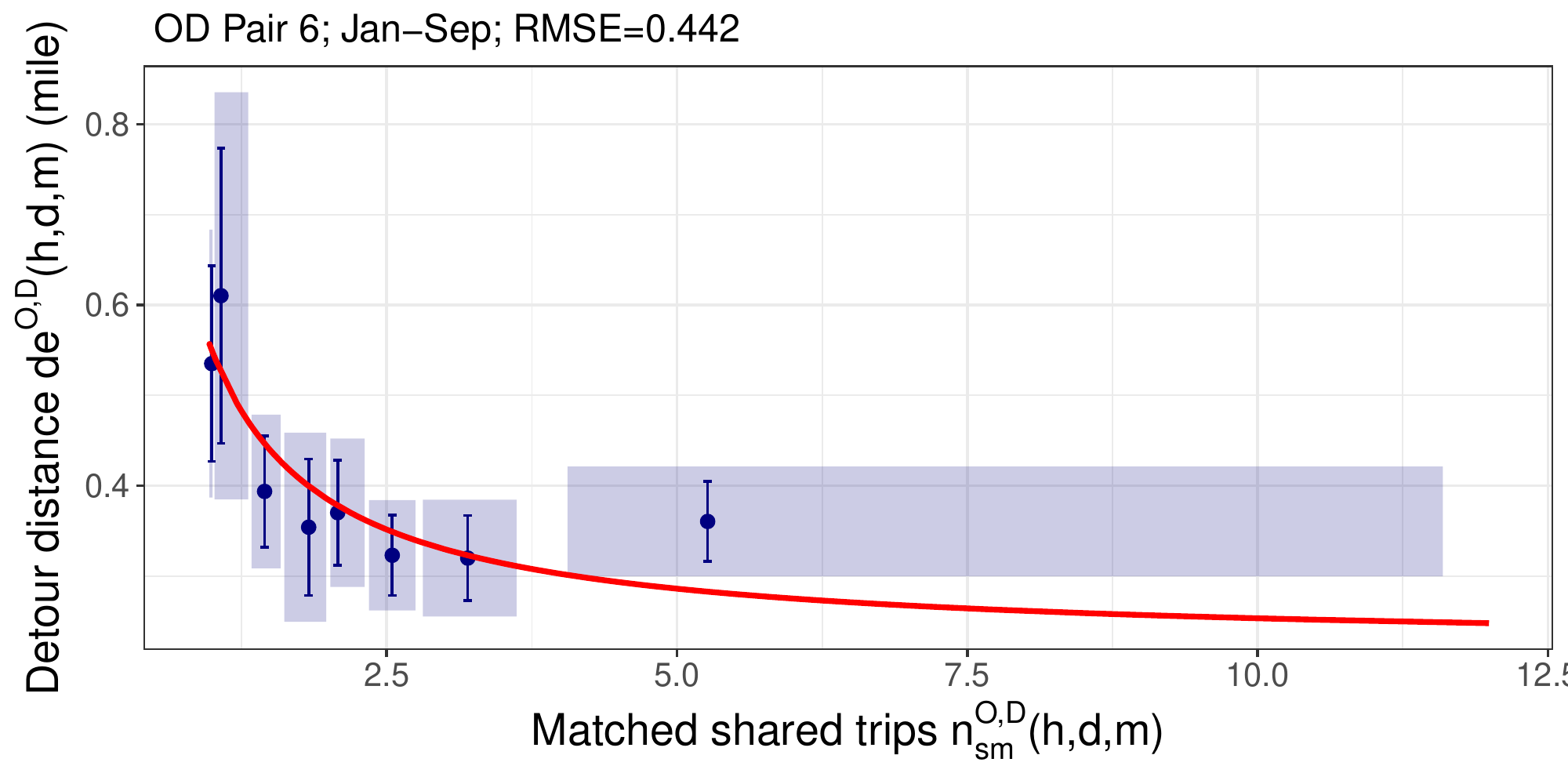}
    \end{subfigure}
    \caption{Detour distance for six OD pairs.}
    \label{fig:detour_dis_3}
\end{figure}

To test the strength and statistical significance of this decreasing trend, we also  fitted regression models for each OD pair. The following functional form was used  based on the trends observed:

\begin{equation}\label{eq:detour_reg}
    de^{O,D}(h,d,m) = \beta_0+\beta_1\frac{1}{n^{O,D}_{s_m}(h,d,m)}.
\end{equation}

The results of the regression are shown in Table \ref{tab:detour_disaggregate}. Note that the purpose of the regression model is to help visualize the trend of the points instead of proposing any causation between the parameters. In addition, the regression is used on the original data, whereas the bins are only used for better visualization. The results of the regression reveal that the decreasing trend is statistically significant across all OD pairs and time periods. 

In addition to the regression results shown in Table \mbox{\ref{tab:detour_disaggregate}}, the elasticity of the regression model \mbox{\eqref{eq:detour_reg}} can be expressed as,

\begin{equation}\label{eq:detour_elasticity}
    \epsilon_{de}^{O,D}(h,d,m)=\frac{dy/y}{dx/x}=-\frac{\beta_1}{\beta_0n^{O,D}_{s_m}(h,d,m)+\beta_1},
\end{equation}
which quantifies the relative change in the dependent variable, i.e., detour distance, based on the relative change in the independent variable, i.e., number of matched shared trips. For the 6 OD pairs shown in Figure \ref{fig:detour_dis_3}, when the number of matched shared trips is 1, an x\% increase in the matched shared trips leads to a 0.499x\%, 0.496x\%, 0.624x\%, 0.613x\%, 0.754x\% and 0.600x\% decrease in the detour distance. The effect becomes less significant when the number of shared trips is large. For example, when the number of matched shared trips is 5, an x\% increase in the matched shared trips leads to a 0.165x\%, 0.163x\%, 0.248x\%, 0.242x\%, 0.371x\% and 0.231x\% decrease in the detour distance. This implies that detour distance becomes relatively insensitive to the number of shared trips when it is large. However, the plots reveal that detour distance can be reduced by a factor of two or more when the number of shared trips increases. 

Note that unlike detour distance, detour time is more complex since it also depends on speed. Figure \ref{fig:per_req} shows a positive relationship between the number of shared trips and the number of total trips. Therefore, a high number of shared trips implies a higher volume at the north side region, which results in lower  travel speeds. As a result, the change of the detour time depends on the relative change in both detour distance and travel speed. In addition, in order to protect personal privacy, the start and end timestamps of all trips are rounded to the nearest 15-minute interval by the TNCs, which makes it impossible to obtain the accurate detour time. Due to the lack of travel speed data and accurate travel time available, this problem is not studied in this paper but is one of our future research directions.

\subsection{Shared trips match rate}

Figure \ref{fig:per_req} shows that the proportion of riders  willing to share rides  increases  with the increase in the number of travelers, at least in the north side region of Chicago in which most trips are occurred. Following that, this section examines if the probability of an authorized shared trip being matched increases as  more riders request  shared trips. The matched percentage for a given hour-day-month combination in region $i$ is defined as: 
\begin{equation}\label{eq:theta_sm}
    \theta_{s_m,i}(h,d,m) = \frac{n_{s_m,i}^t(h,d,m)}{n_{s_a,i}^t(h,d,m)}\times 100 \quad \text{(\% matched shared trips)}.
\end{equation}

\begin{figure}[htb]
    \centering
    \includegraphics[width=4in]{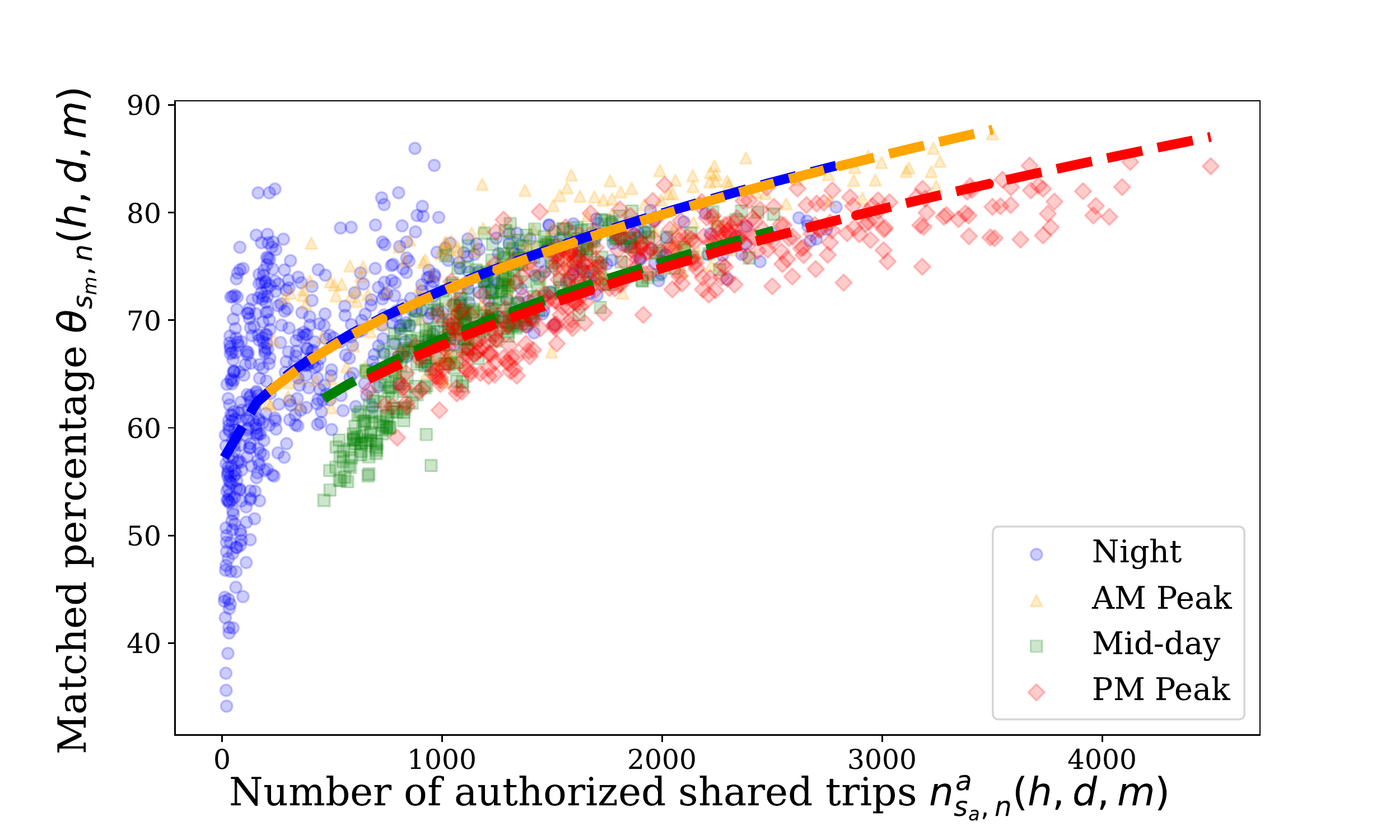}
    \caption{Matched percentage of shared trips in the north region, RMSE=4.891}
    \label{fig:acceptance_north}
\end{figure}

\begin{figure}[htb]
    \centering
    \begin{subfigure}{0.45\textwidth}
        \centering
        \includegraphics[width=\textwidth]{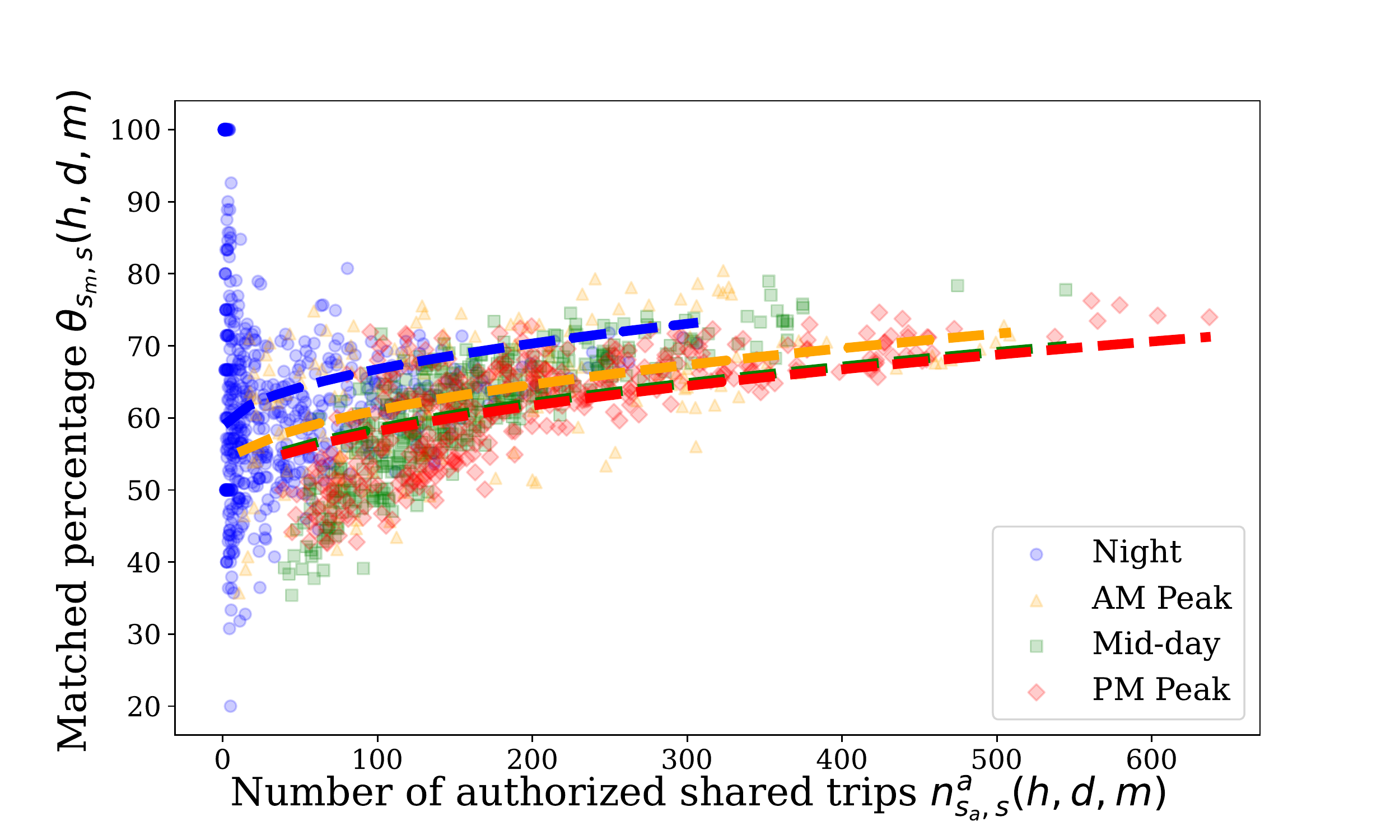}
        \caption{Before filtering, RMSE=10.199}
        \label{fig:acceptance_south_before_filtering}
    \end{subfigure}
    \begin{subfigure}{0.45\textwidth}
        \centering
        \includegraphics[width=\textwidth]{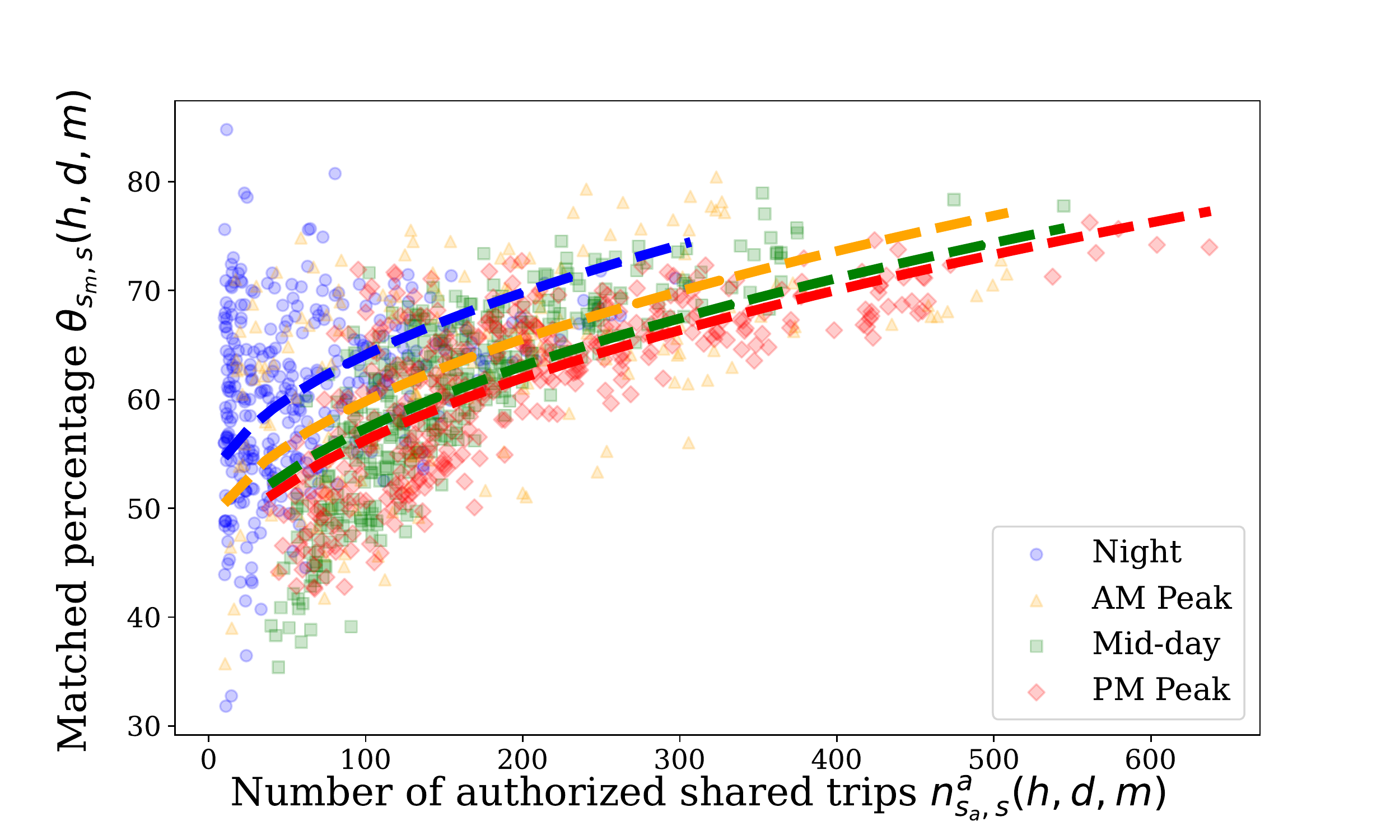}
        \caption{After filtering, RMSE=6.548}
        \label{fig:acceptance_south_after_filtering}
    \end{subfigure}
    \caption{Matched percentage of shared trips in the south region.}
    \label{fig:acceptance_south}
\end{figure}

Figures \ref{fig:acceptance_north}-\ref{fig:acceptance_south} show the relationship between the matched percentage and the number of authorized shared trips for both regions. We divided the data into the four time-of-day periods defined in Figure \ref{fig:per_req}. Figures \ref{fig:acceptance_north}-\ref{fig:acceptance_south} reveal that matched percentage rises with the number of authorized shared trips for both regions, which is in line with our expectation for economies of scale. When the percentage of authorized shared trips in the network rises, more authorized trips likely start and end in near locations where other trips are authorized, and consequently, more requests are matched. However, the increasing trend observed in the south side region is not as significant as what is observed in the north side. This may be  because there are points with extremely low number of authorized matched trips in the night hours that are matched. As a result, the matched percentage is extremely high, e.g., 100\%, for these points as shown in \ref{fig:acceptance_south_before_filtering}. After we removed the points with less than 10 authorized shared trips in that region (213 points are filtered out), we obtained the results shown in \ref{fig:acceptance_south_after_filtering}, which shows a more significant increase in the matched percentage associated with the increase in the authorized shared trips. 

A period-wise regression model in the following form was fitted based on the visual  patterns observed in Figures \mbox{\ref{fig:acceptance_north}-\ref{fig:acceptance_south}},

\begin{table}[!htbp] \centering 
  \caption{Regression results for detour distance (mile)} 
  \label{tab:detour_disaggregate} 
\begin{tabular}{@{\extracolsep{5pt}}lcccccc} 
\\[-1.8ex]\hline 
\hline \\[-1.8ex] 
 & \multicolumn{6}{c}{\textit{Dependent variable:}} \\ 
\cline{2-7} 
\\[-1.8ex] & \multicolumn{6}{c}{Detour distance $de^{O,D}(h,d,m)$} \\ 
\\[-1.8ex] & (1) & (2) & (3) & (4) & (5) & (6)\\ 
\hline \\[-1.8ex] 
 $\frac{1}{n^{O,D}_{s_m}(h,d,m)}$ & 0.341$^{***}$ & 0.234$^{***}$ & 0.380$^{***}$ & 0.424$^{***}$ & 0.431$^{***}$ & 0.330$^{***}$ \\ 
  & (0.060) & (0.047) & (0.058) & (0.057) & (0.122) & (0.056) \\ 
  Constant & 0.348$^{***}$ & 0.238$^{***}$ & 0.231$^{***}$ & 0.266$^{***}$ & 0.145$^{*}$ & 0.219$^{***}$ \\ 
  & (0.036) & (0.031) & (0.040) & (0.037) & (0.083) & (0.036) \\ 
 \hline \\[-1.8ex] 
R$^{2}$ & 0.033 & 0.029 & 0.048 & 0.059 & 0.014 & 0.040 \\ 
Adjusted R$^{2}$ & 0.032 & 0.028 & 0.047 & 0.058 & 0.013 & 0.039 \\ 
\hline 
\hline \\[-1.8ex] 
Significance levels & \multicolumn{6}{r}{$^{*}$p$<$0.1; $^{**}$p$<$0.05; $^{***}$p$<$0.01} \\ 
\end{tabular} 
\end{table} 

\begin{equation}\label{eq:regression_mr}
    \theta_{s_m,i}(h,d,m) = \beta_0+\beta_1\sqrt{n^{a}_{s_a,i}(h,d,m)}+\delta^h_A\beta_{AM Peak}+\delta^h_M\beta_{Mid-day}+\delta^h_P\beta_{PM Peak}.
\end{equation}

In Equation \eqref{eq:regression_mr}, the coefficient $\beta_1$ is fixed for all periods by regarding the period as a categorical parameter and using ``Night'' hours as the reference level, and $\delta^h_{i}$'s are binary variables indicating the period of the data point. The results of the regression are shown in Table \ref{tab:matching_rate_revised}. Note that the coefficients of $\sqrt{n^a_{s_a,i}(h,d,m)}$ for all models are positive and significant, which represent a statistically significant positive relationship between matched rate and number of authored shared trips. In addition, the elasticity for the regression model \eqref{eq:regression_mr} can be expressed as,

\begin{equation}\label{eq:mr_elasticity}
    \epsilon_{\theta_{s_m,i}}(h,d,m)=\frac{dy/y}{dx/x}=\frac{\beta_1}{2}\frac{\sqrt{x}}{\beta_{0}+\beta_1\sqrt{x}+\delta^h_A\beta_{AM Peak}+\delta^h_M\beta_{Mid-day}+\delta^h_P\beta_{PM Peak}},
\end{equation}

Unlike the detour distance, the overall matched percentage does not present the flatness within the range of the authorized shared trips in the dataset. For example, for the PM Peak period, when $n_{s_a}^a$ increases from 1,000 to 4,000 vph in the north region, the elasticity increases from 0.128 to 0.203; in the south region, when $n_{s_a}^a$ after filtering increases from 100 to 600 vph, the elasticity increases from 0.122 to 0.221. This indicates that although the south side has lower shared trips than the north side, the relative effect of authorized shared trips on the matched percentage in the south side is more significant than the north side.

\begin{table}[!htbp] \centering 
  \caption{Regression results for matched percentage of shared trips} 
  \label{tab:matching_rate_revised} 
\begin{tabular}{@{\extracolsep{5pt}}lcccccc} 
\\[-1.8ex]\hline 
\hline \\[-1.8ex] 
 & \multicolumn{3}{c}{\textit{Dependent variable:}} \\ 
\cline{2-4} 
\\[-1.8ex] & \multicolumn{3}{c}{Matched percentage $\theta_{s_m,i}(h,d,m)$} \\ 
\\[-1.8ex] & N Jan-Sep & S Jan-Sep before & S Jan-Sep after\\ 
\hline \\[-1.8ex] 
 $\sqrt{n_{s_a,i}^a(h,d,m)}$ & 0.546$^{***}$ & 0.859$^{***}$ & 1.378$^{***}$ \\ 
  & (0.013) & (0.071) & (0.051) \\ 
  AM Peak & $-$0.120 & $-$5.704$^{***}$ & $-$4.255$^{***}$ \\ 
  & (0.455) & (0.996) & (0.646) \\ 
  Mid-day & $-$4.535$^{***}$ & $-$8.234$^{***}$ & $-$6.745$^{***}$ \\ 
  & (0.362) & (0.843) & (0.549) \\ 
  PM Peak & $-$5.083$^{***}$ & $-$8.621$^{***}$ & $-$7.831$^{***}$ \\ 
  & (0.429) & (0.904) & (0.584) \\ 
  Constant & 55.515$^{***}$ & 58.172$^{***}$ & 50.326$^{***}$ \\ 
  & (0.335) & (0.573) & (0.512) \\ 
 \hline \\[-1.8ex] 
R$^{2}$ & 0.631 & 0.096 & 0.372 \\ 
Adjusted R$^{2}$ & 0.631 & 0.094 & 0.370 \\ 
\hline 
\hline \\[-1.8ex] 
Significance levels & \multicolumn{3}{r}{$^{*}$p$<$0.1; $^{**}$p$<$0.05; $^{***}$p$<$0.01} \\ 
\end{tabular} 
\end{table}

The analysis by time-period shows that  points from the AM peak, Mid day and PM peak time periods are tighter and closer to the regression line than points during the Night time period. One possible reason for this phenomenon is that people may tend to avoid travel with strangers in the same car in night hours for safety reasons. Another is that the distribution of the OD pairs in the night hours may be more random and scattered than others. Therefore, the match rate is also more random.

\subsection{Unit fare ratios}

\begin{figure}[htb]
    \centering
    \includegraphics[width=4in]{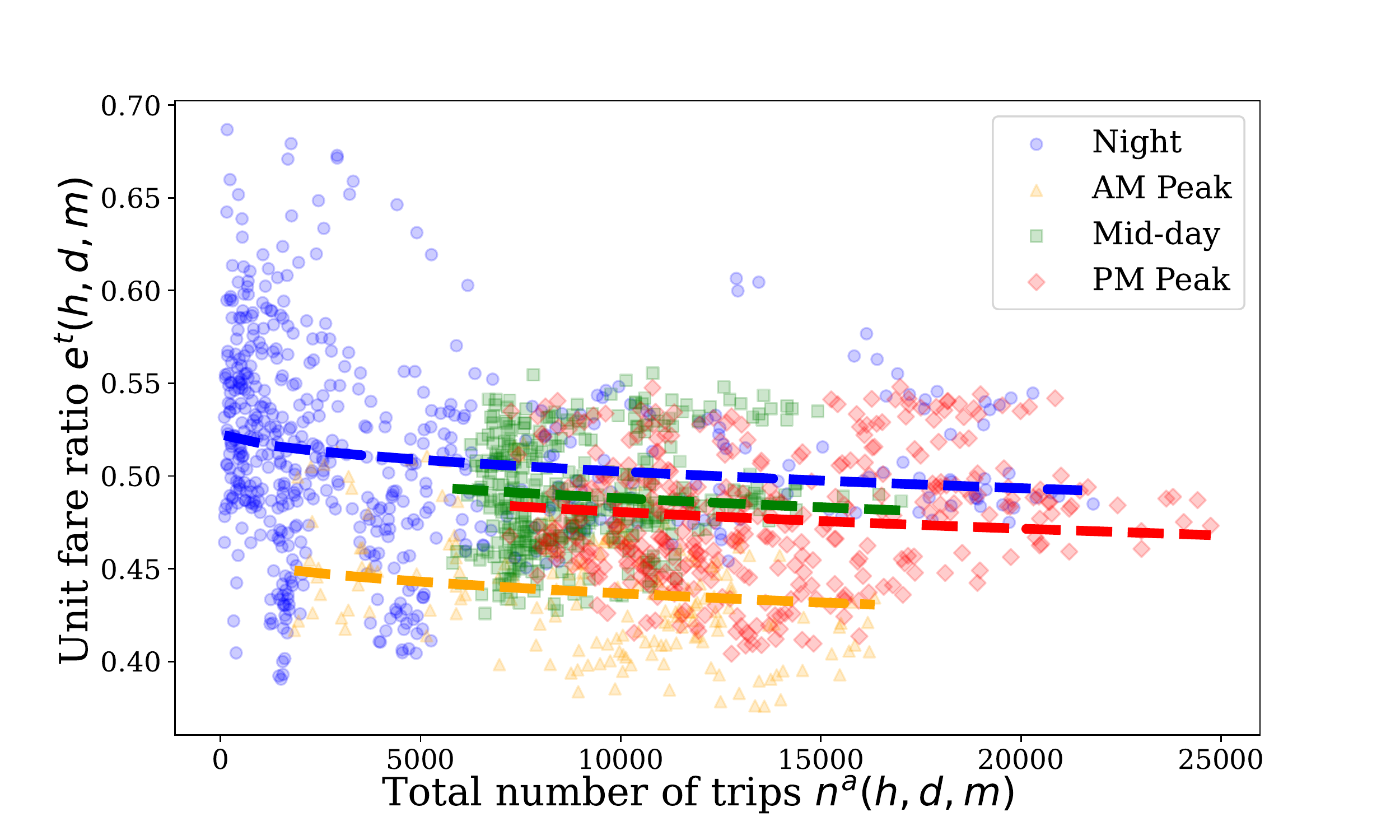}
    \caption{Time-based unit fare ratio, RMSE=0.041}
    \label{fig:cost_ratio_time}
\end{figure}

\begin{figure}[htb]
    \centering
    \includegraphics[width=4in]{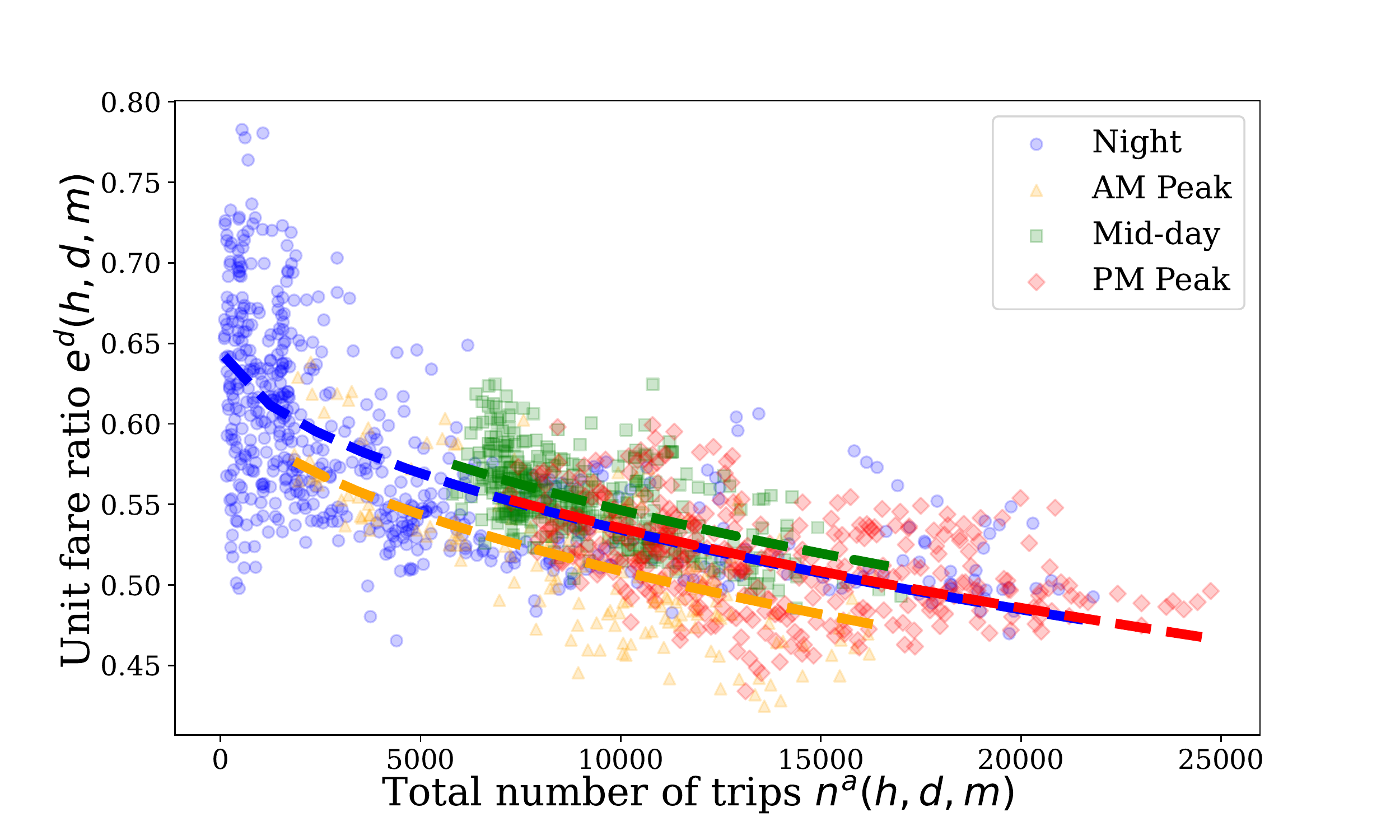}
    \caption{Distance-based unit fare ratio, RMSE=0.036}
    \label{fig:cost_ratio_dis}
\end{figure}

We now consider two normalized metrics of monetary trip cost. Unit fares per distance (\$/mi) and time (\$/min) for each trip were obtained by dividing the total fare by the trip distance and time, respectively. Note that in order to protect personal privacy, the fare for a trip is rounded to the nearest \$2.50, and the tip for a trip is rounded to the nearest \$1.00, which is a limitation for this analysis. The ratio of the unit fare between shared trips and single trips is referred to as the fare ratio. Figure \ref{fig:cost_ratio_time} and Figure \ref{fig:cost_ratio_dis} show the time-based and distance-based fare ratio, respectively, as a function of the total number of trips. As expected, the unit fare ratios are always less than 1, which means that costs of shared trips are always lower than  single trips. For all periods, the unit fare ratio decreases with the increase of volume, and the coefficients are all significant. These results could imply that as traffic volume increases, the travelers are encouraged to share trips to mitigate the traffic burden. The same regression model as Equation \mbox{\eqref{eq:regression_mr}} was fitted, and Table \mbox{\ref{tab:cost_ratio_revised}} provide the regression results that demonstrate these findings are statistically significant. In addition, when the number of trips increases from 10,000 to 20,000, the elasticity for the distance-based ratio decreases from -0.112 to -0.174, and the elasticity for the time-based ratio decreases from -0.022 to -0.032. This indicates that traffic volumes have a more significant impact on distance-based unit ratio. As mentioned in Section \ref{sec:detour}, although it is reasonable to assume that the traffic volume would not impact the travel distance, the influence on travel time cannot be ignored. With the increase in traffic volume, travel time is expected to be increased. As a result, assuming the total fare does not change, the time-based unit fares for both modes are decreased. Consequently, for a fixed reduction in the total fare, the reduction in the time-based unit fare ratio is less than the distance-based unit fare ratio.

\begin{table}[!htbp] \centering 
  \caption{Regression results for cost ratio} 
  \label{tab:cost_ratio_revised} 
\begin{tabular}{@{\extracolsep{5pt}}lcccc} 
\\[-1.8ex]\hline 
\hline \\[-1.8ex] 
 & \multicolumn{2}{c}{\textit{Dependent variable:}} \\ 
\cline{2-3} 
\\[-1.8ex] & Time-based $e^t$ & Distance-based $e^d$ \\ 
\hline \\[-1.8ex] 
 $\sqrt{n^a(h,d,m)}$ & $-$0.0002$^{***}$ & $-$0.001$^{***}$ \\ 
  & (0.00004) & (0.00004) \\ 
  AM Peak & $-$0.066$^{***}$ & $-$0.026$^{***}$ \\ 
  & (0.004) & (0.003) \\ 
  Mid-day & $-$0.014$^{***}$ & 0.012$^{***}$ \\ 
  & (0.003) & (0.003) \\ 
  PM Peak & $-$0.022$^{***}$ & 0.001 \\ 
  & (0.004) & (0.003) \\ 
  Constant & 0.524$^{***}$ & 0.654$^{***}$ \\ 
  & (0.003) & (0.003) \\ 
 \hline \\[-1.8ex] 
R$^{2}$ & 0.256 & 0.571 \\ 
Adjusted R$^{2}$ & 0.254 & 0.570 \\ 
\hline 
\hline \\[-1.8ex] 
Significance levels & \multicolumn{2}{r}{$^{*}$p$<$0.1; $^{**}$p$<$0.05; $^{***}$p$<$0.01} \\ 
\end{tabular} 
\end{table}

\section{Concluding remarks}\label{sec:conclusions}
This paper demonstrates evidence of economies of scale and increasing returns to scale in ridesplitting services using empirical data from 2019 reported to the City of Chicago. Due to the suspicious phenomena observed in the data between October-December, this period is excluded from this study. Economies of scale manifests as detour lengths decreasing as the number of ridesplitting requests increases, while increasing returns to scale manifests as increased rates of trips being matched. Specifically, when the number of hourly shared trips between an OD pair is 1, the average elasticity of the detour distance of the 16 OD pairs with the highest number of trips is -0.584. The relative effect on detour distance decreases with the increase in the number of shared trips, e.g., when the number of hourly shared trips between an OD increases to 5, the corresponding average elasticity changes to -0.226. Two other knock-on effects are also observed: as the number of authorized shared trips rises, the willingness-to-share increases and the unit fare ratio between the shared trips and single trips declines. Moreover, the WTS on weekends is lower than weekdays although the total number of trips is higher on weekends. For example, the average willingness-to-share decreases from 17.3\% on weekdays to 13.6\% on weekends in the north side region. The elasticity for the distance-based unit fare ratio is higher than the time-based unit fare ratio due to the increase in travel time with traffic volumes. Note that although the number of authorized shared trips is the independent variable in all theories above, we use other alternatives, such as number of matched shared trips and total number of shared trips, in order to make the expressed relationship natural. For example, Figure \ref{fig:detour_dis_3} and Figure \ref{fig:detour_rest_10} use the number of matched shared trips as the independent variable since it is straightforward to show the relationship between the detour distance and the number of matched shared trips rather than the number of authorized shared trips. Nevertheless, all these alternatives are positively related to the number of authorized shared trips, so they support the conclusions we made. Overall, the findings in this paper provide the existence of benefits from the rise in the usage of ridesplitting service. This could lead to a positive feedback loop posited in \citet{Lehe2020}, in which increasing ridesplitting usage should improve the quality of the service, resulting in even more incentive for riders to use the service. Thus, it suggests efforts and strategies should be made to encourage and enhance the ridesplitting service to leverage this positive feedback mechanism brought about by the economies of scale and increasing returns to scale. 

While the empirical trends observed  are indicative of the aforementioned economies of scales and increasing returns to scale phenomena, they alone cannot confirm their presence in ridesplitting. It is still possible that there are other reasons that these trends are observed. Still, the observed trends are in line with what we would expect given that these phenomena were to occur and compelling enough to attribute to these reasons. Future work should endeavor to isolate these effects, identify causation and confirm their existence in other datasets.

Compared to detour distance, detour time is a more complex variable since it relies on both detour distance and travel speed, on which the travel demand has opposite influences. The impact of ridesplitting on detour time could be a promising research direction.

Despite the economies of scale and increasing returns to scale in ridesplitting observed in this paper, there is debate on how TNC trips impact traffic congestion. Some studies \citep{rayle2016just,clewlow2017disruptive,zimmer2016third, feigon2016shared,feigon2018broadening} have argued that ridesplitting is likely to reduce traffic congestion; however, \citet{roy2020traffic} and \citet{erhardt2019transportation} revealed that TNCs are associated with the increased congestion in San Francisco. \citet{erhardt2022has} claimed that TNCs account for around 10\% reduction in bus and rail ridership in mid-sized metropolitan areas by 2018. \citet{schaller2021can} also showed that the ride-hail service can increase vehicle miles traveled (VMT) considerably due to the addition of dead-head miles and the user shift from public transportation to ride-hail service, so it was suggested to prioritize public transportation over ride-hail service in dense urban centers.  Therefore, it is of interest to investigate the effect of ridesplitting on the traffic operations in the City of Chicago.

Furthermore, due to the COVID-19 crisis, the ridesplitting service was ceased in March 2020, and it only returned to Chicago in June 2022. Lyft and Uber pooling returned to New York City in August 2021 and June 2022, respectively. It is a promising topic to investigate if similar scale effects exist after the re-introduction of ridesplitting and the impact of the pandemic on ridesplitting patterns.

\section*{Acknowledgement}
This research was supported by NSF Grants CMMI-2052337 and CMMI-2052512.

\appendix
\section*{Appendices}

\addcontentsline{toc}{section}{Appendices}
\renewcommand{\thesubsection}{\Alph{subsection}}
\renewcommand\thefigure{\arabic{figure}}
\renewcommand\thetable{\arabic{table}}
\subsection{Systematic differences between January-September and October-December 2019}\label{sec:a1}
This Appendix shows the systematic differences between the period January-September and the period October-December 2019 and the unexpected patterns existing in both periods.

\subsubsection*{Number of shared trips}

\begin{figure}[htb]
    \centering
    \includegraphics[width=4.5in]{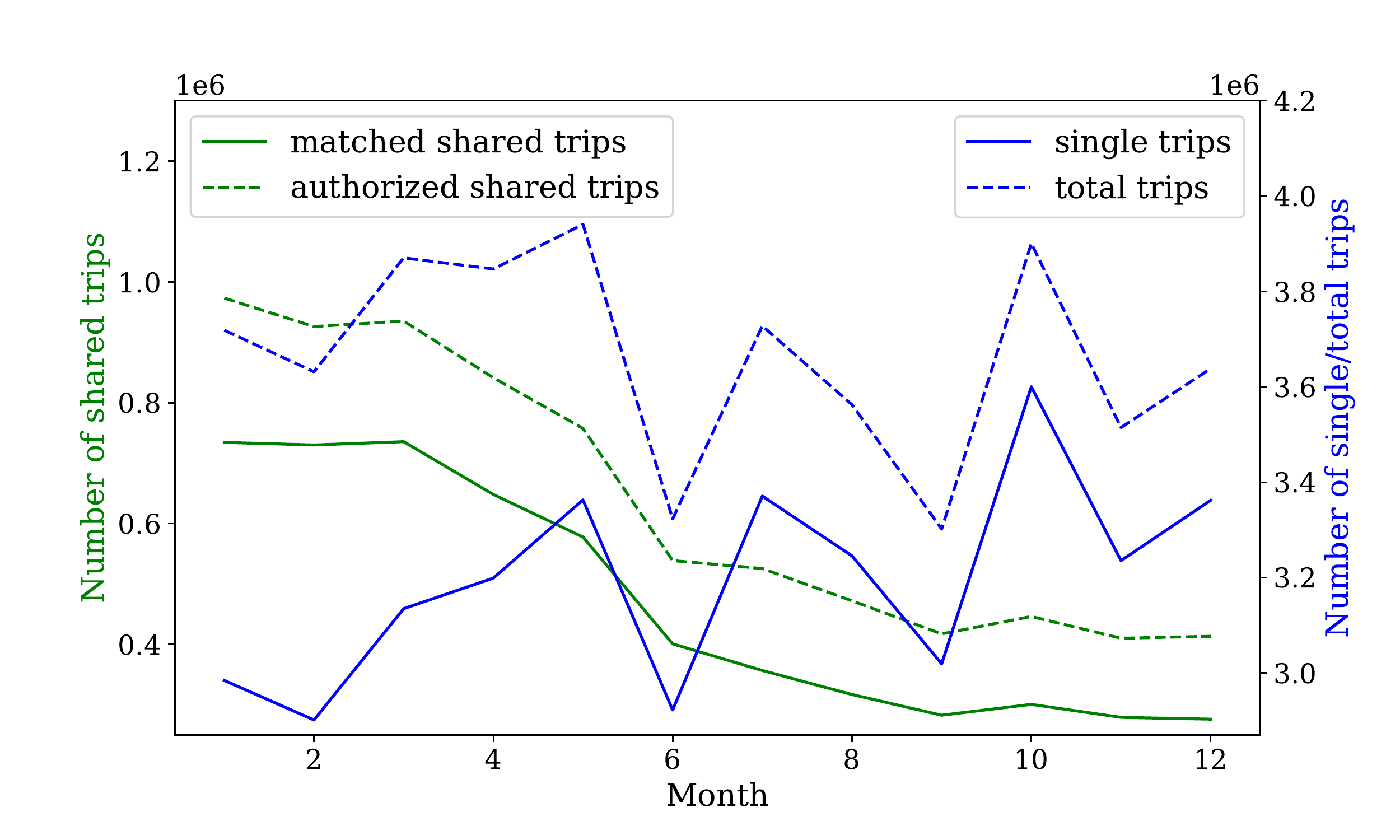}
    \caption{Change of number of trips with month.}
    \label{fig:n_trip_month}
\end{figure}

Figure \ref{fig:n_trip_month} shows the numbers of single trips, shared trips and total trips in each month in 2019. While the number of single trips is relatively stable across the year, the number of authorized and matched shared trips steadily decreased from January through September, and they become stable after October. \citet{taiebat2022sharing} found the same pattern and claimed that one possible reason for the continuous decline in  the number of shared trips before September was the increasing total cost of a shared trip. However, they also showed that the unit cost (per distance) of a shared trip was  stable before September 2019 and decreased only after October 2019.

\subsubsection*{Travel and detour distance}
\begin{figure}[htb]
    \centering
    \begin{subfigure}{0.45\textwidth}
        \centering
        \includegraphics[width=\textwidth]{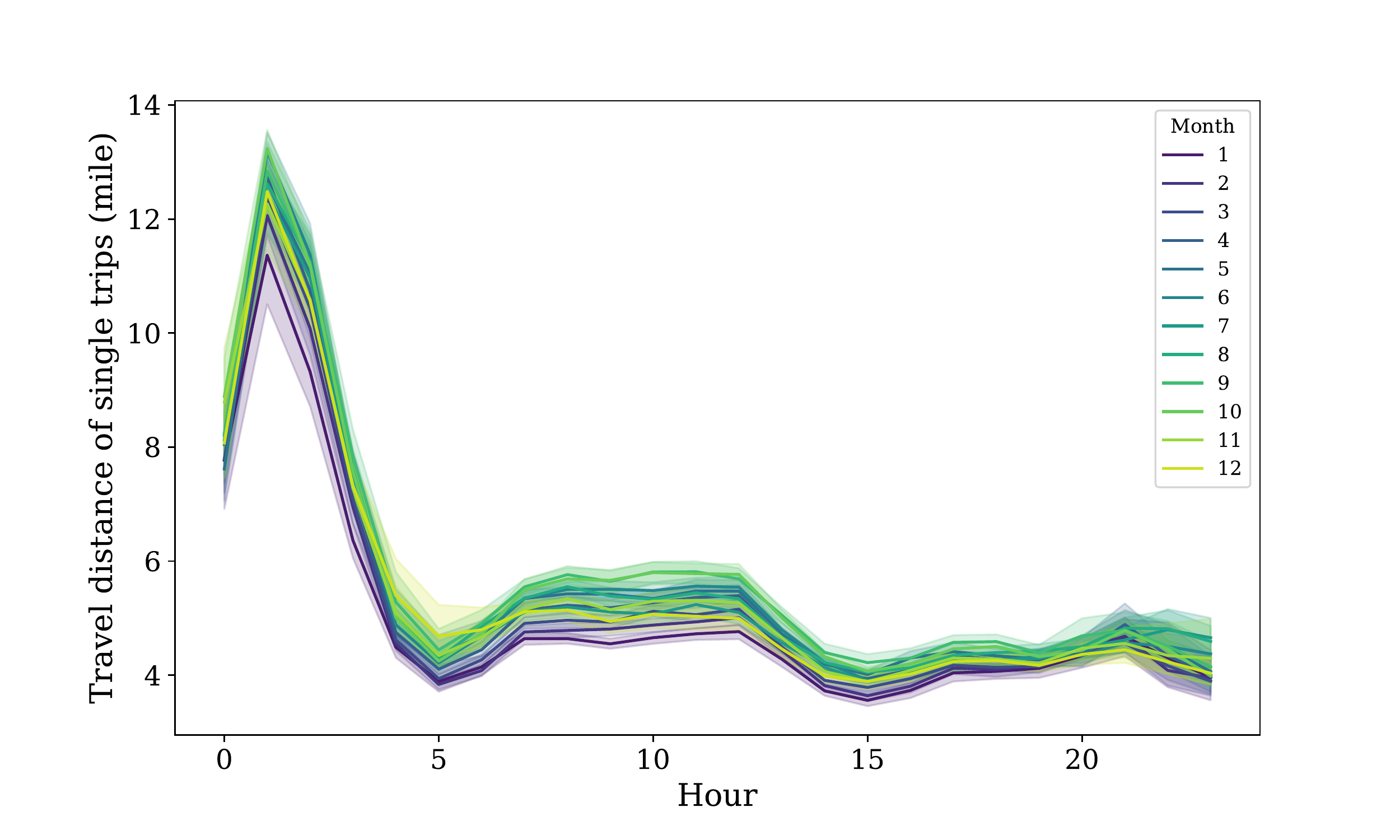}
        \caption{Single trips.}
        \label{fig:dis_single}
    \end{subfigure}
    \begin{subfigure}{0.45\textwidth}
        \centering
        \includegraphics[width=\textwidth]{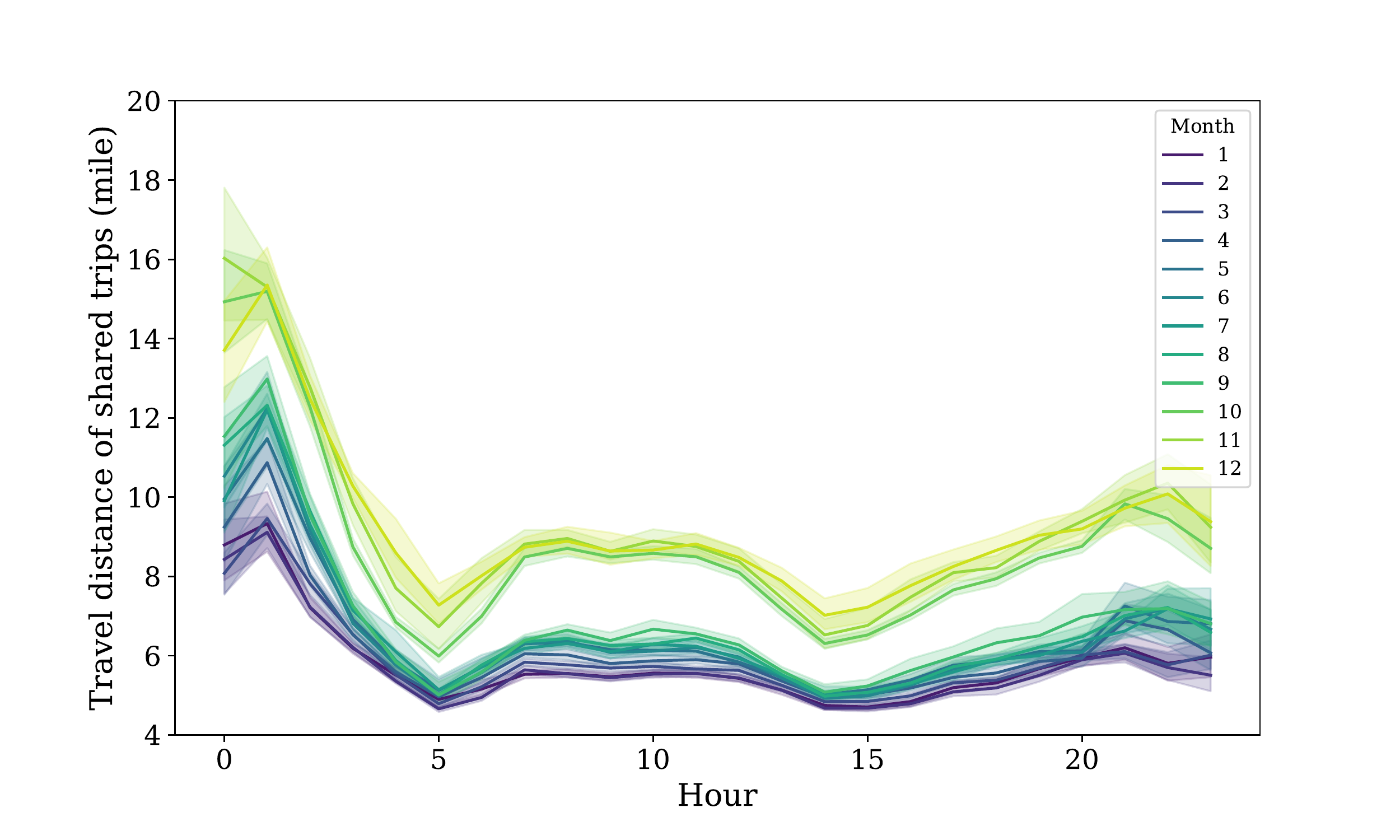}
        \caption{Shared trips.}
        \label{fig:dis_shared}
    \end{subfigure}
    \caption{Trip length.}
    \label{fig:trip_dis}
\end{figure}

In addition to the volumes, travel distances also differ for shared trips between those two periods. Figure \ref{fig:trip_dis} shows how the average trip length varies across a given hour in the day for each of the months in 2019 for both single and shared trips. Notice that travel distance for single trips is consistent across all months while the travel distance for shared trips exhibits a significant increase between the two periods, even though travel distances for shared trips are relatively stable within each of these periods.  

\begin{figure}[h!]
    \centering
    \includegraphics[width=4.5in]{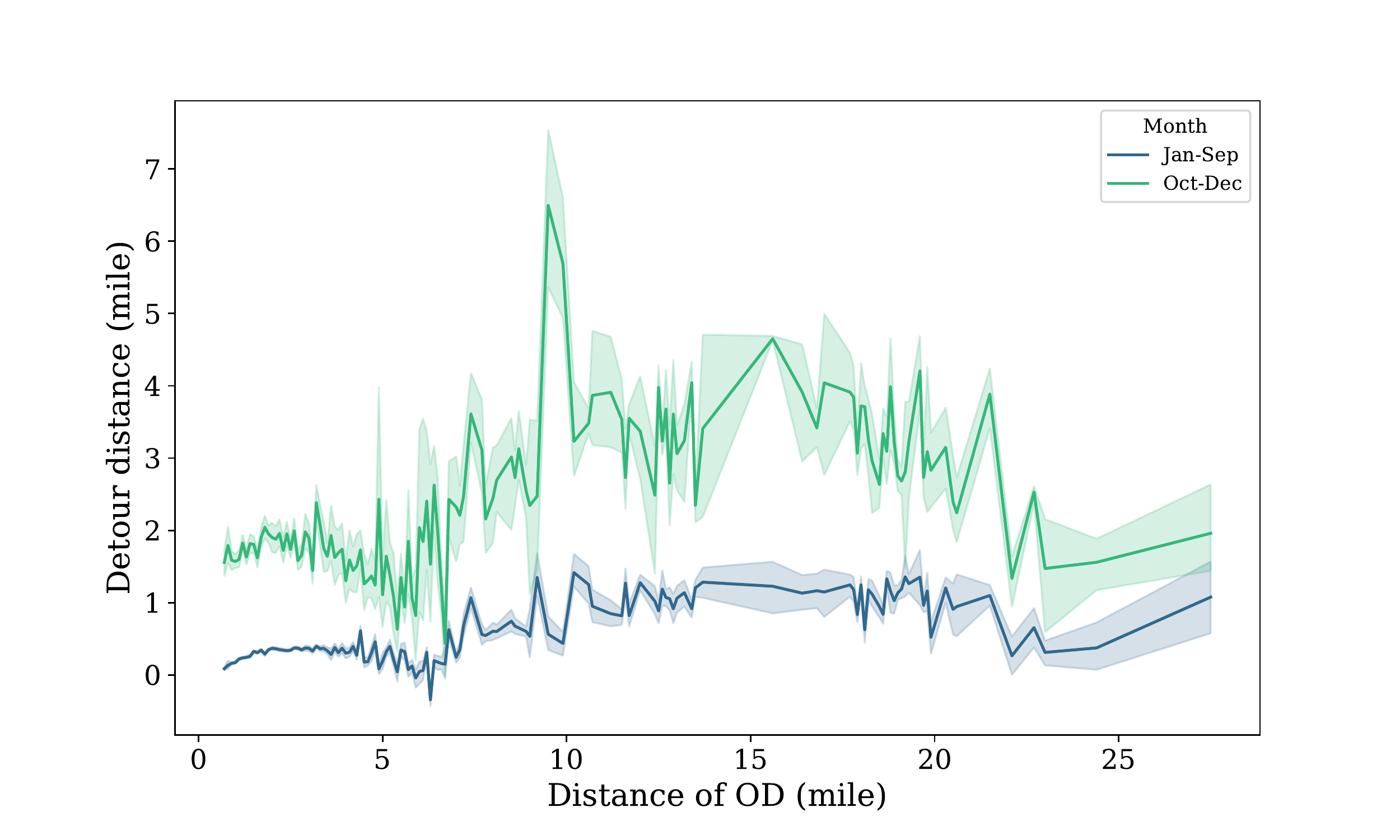}
    \caption{Detour distance.}
    \label{fig:detour_month}
\end{figure}

Does the difference in the travel distance mean people tend to pool for longer trips after October? To test this, Figure \ref{fig:detour_month} shows the detour distance compared to the actual travel distance between the pickup and dropoff OD pairs. According to Figure \ref{fig:dis_single}, the length for single trips in 2019 is stable so we believe this is an appropriate estimation for the distance of ODs. Then, we grouped the data into two classes: January-September and October-December. Figure \ref{fig:detour_month} shows that for a fixed distance between OD pairs, the detour distance in January-September is significantly lower than October-December. This suggests that the increase in shared trip length shown in Figure \ref{fig:dis_shared} results from the rise in detour distance rather than the distance between the origin and destination. In addition, the TNCs have switched from picking up and dropping off customers at their actual origins and destinations to having customers walk to a pick-up point and then walk from a drop-off point in some cities, which should not increase the detour distances. Moreover, Figure \ref{fig:detour_month} shows that, the detour distance for OD pairs shorter than 2 miles is between 1.5-2 miles, which is higher than expected. This excessive increase in the detour distance makes the travel distance data in October-December suspicious.

\subsubsection*{Travel speed}

\begin{figure}[h!]
    \centering
    \includegraphics[width=4.5in]{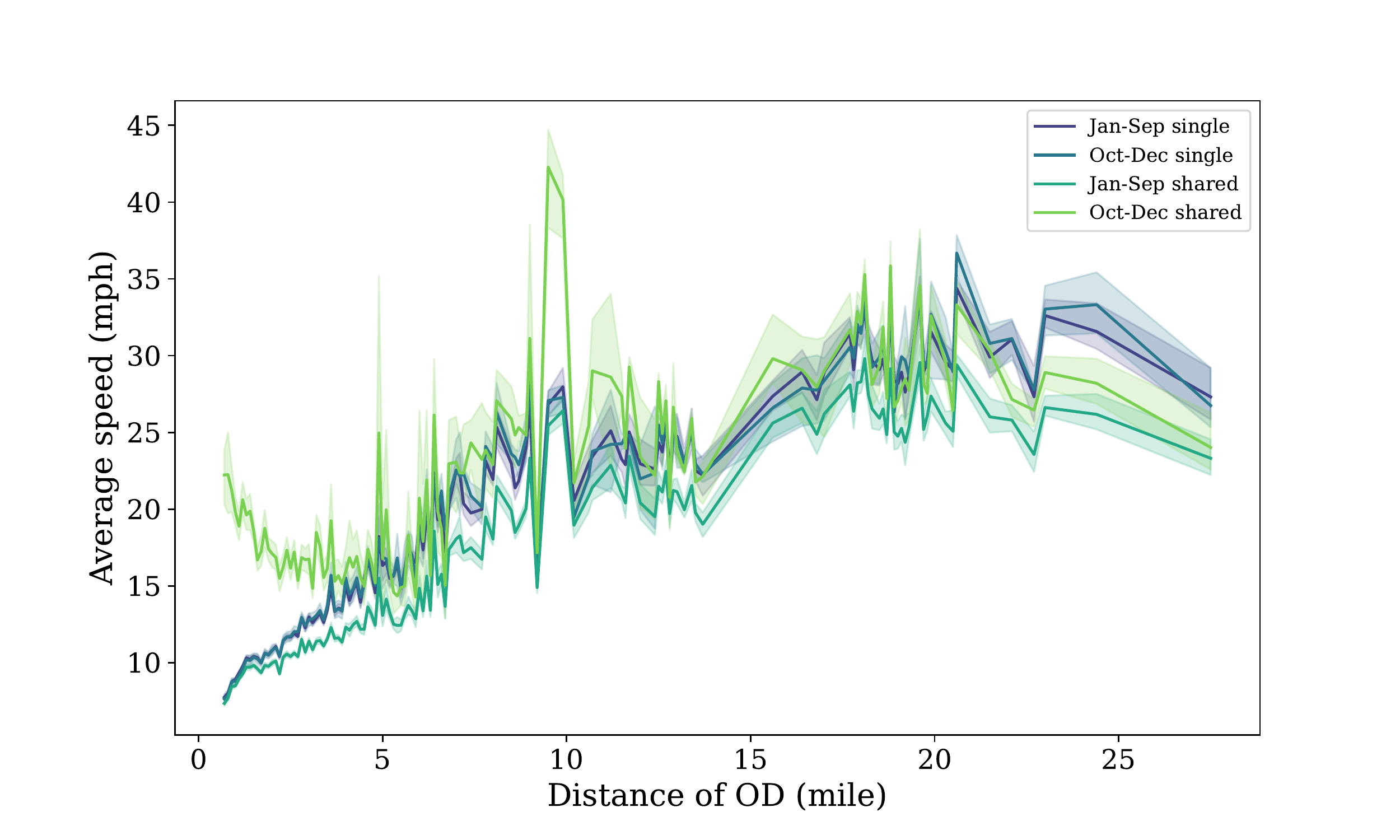}
    \caption{Speed comparison.}
    \label{fig:speed_month}
\end{figure}

Similar to Figure \ref{fig:detour_month}, Figure \ref{fig:speed_month} shows the average travel speed for both single and shared trips associated with the distance between OD pairs. The average speed for single trips between those two periods is very similar. In January-September, the speed for shared trips is lower than the speed for single trips, which is in line with our expectation. However, in October-December, the speed for shared trips is higher, which is another erratic pattern existing between October-December.

Unfortunately, there is no evidence by which to explain these phenomena. The unexpected patterns observed in the detour distance and travel speed between October-December make the data in this period suspicious. To avoid confusion, this period is excluded from this paper\footnote{Despite the unexpected phenomena observed in the data between October-December 2019, we confirmed that all findings in this paper hold in this period as well.}.

\subsection{Detour distance for additional 10 OD pairs}\label{sec:a2}
Figure \ref{fig:detour_rest_10} shows the detour distance for the rest 10 of the 16 OD pairs with the highest number of trips in January-September 2019.
\begin{figure}[H]
    \centering
    \begin{subfigure}{0.45\textwidth}
        \centering
        \includegraphics[width=\textwidth]{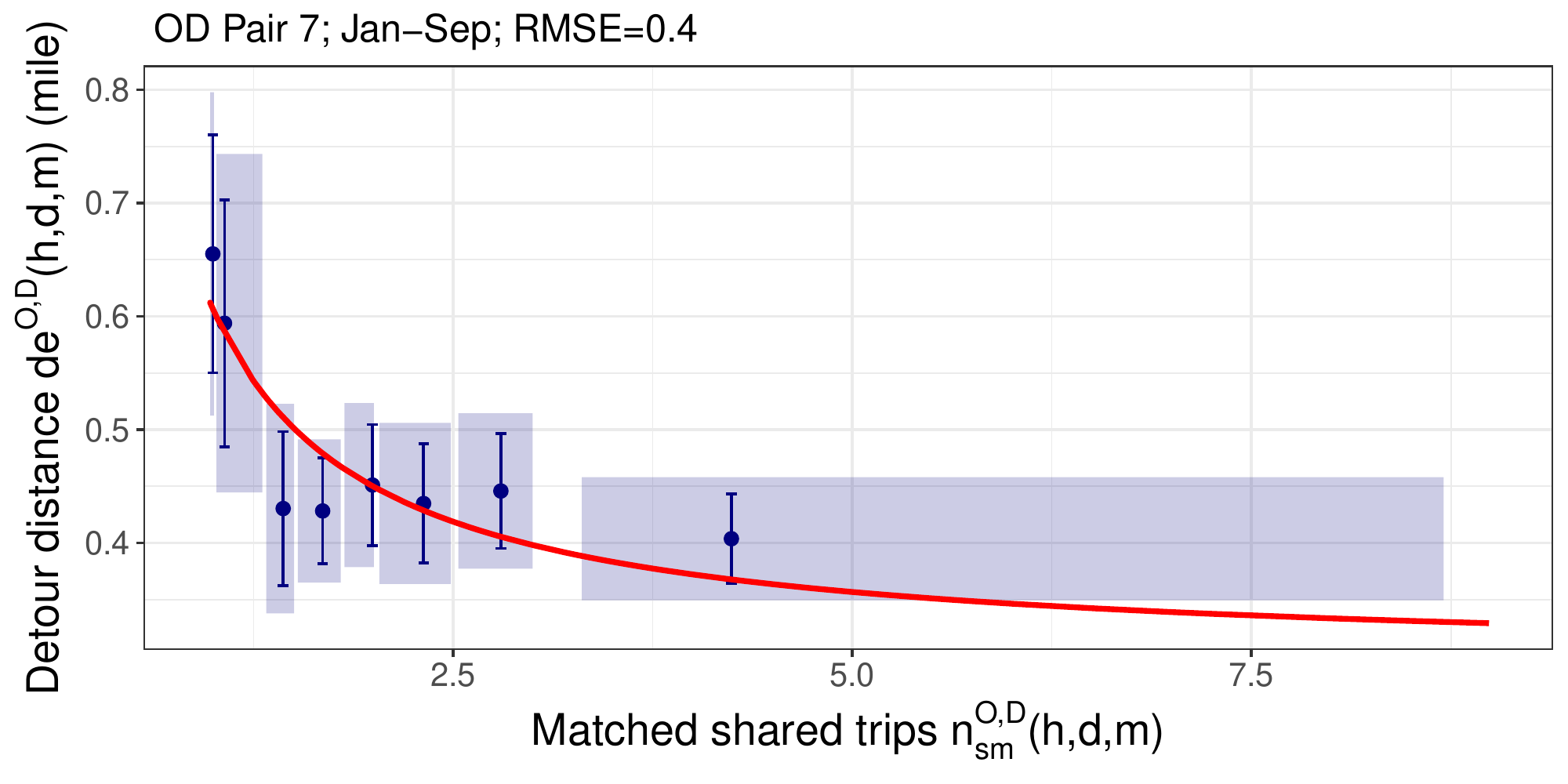}
    \end{subfigure}
    \begin{subfigure}{0.45\textwidth}
        \centering
        \includegraphics[width=\textwidth]{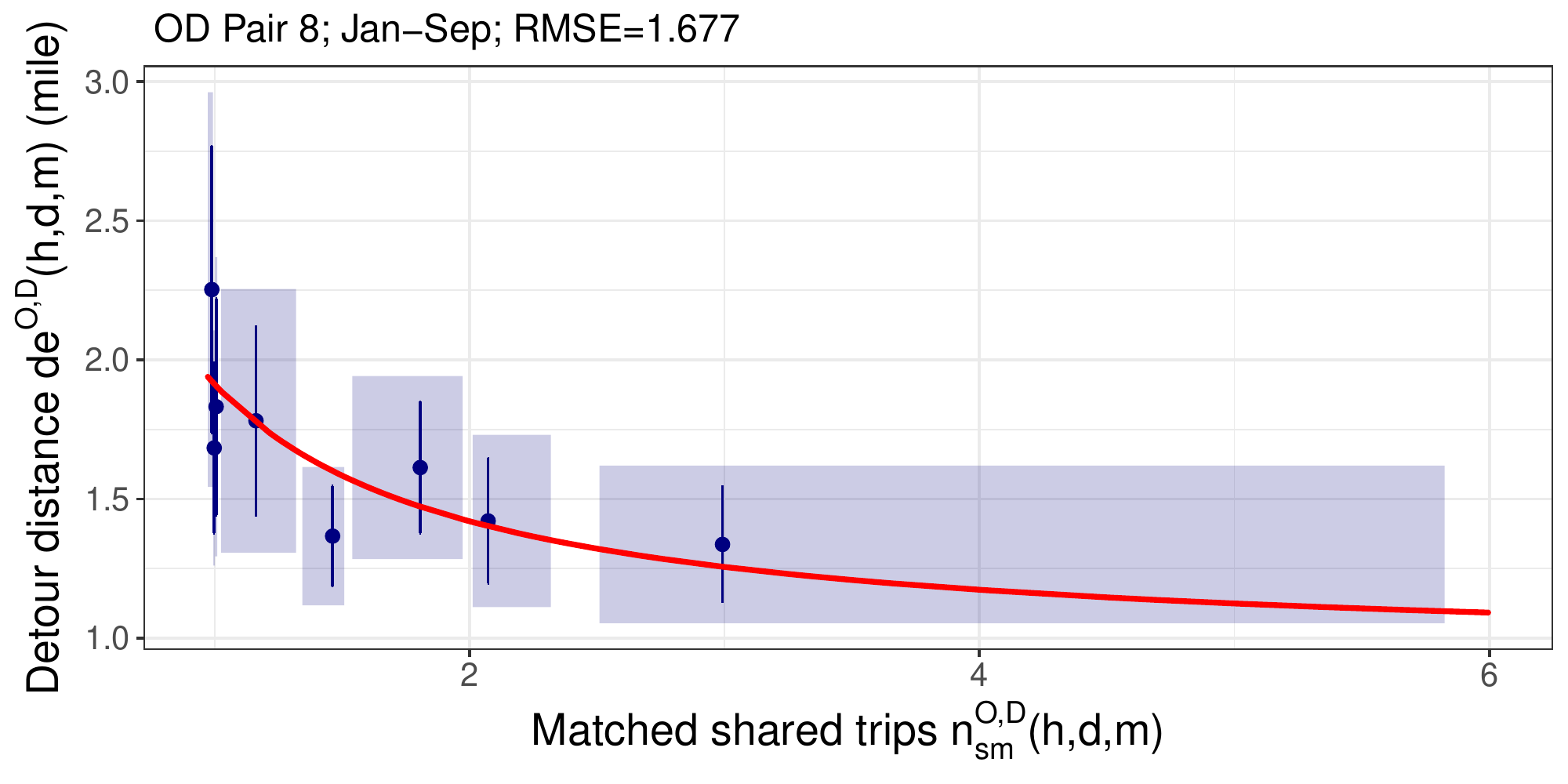}
    \end{subfigure}

    \begin{subfigure}{0.45\textwidth}
        \centering
        \includegraphics[width=\textwidth]{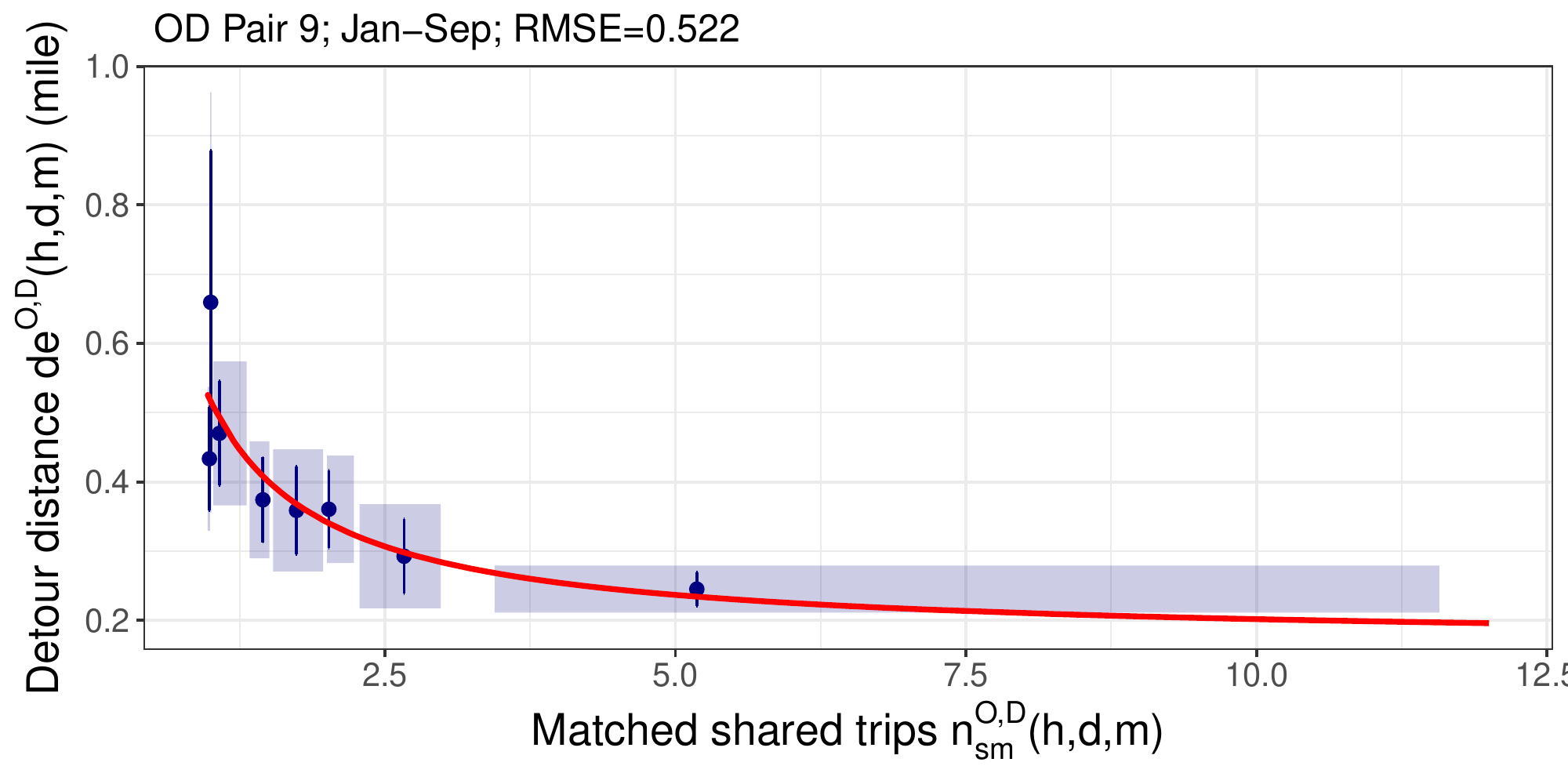}
    \end{subfigure}
    \begin{subfigure}{0.45\textwidth}
        \centering
        \includegraphics[width=\textwidth]{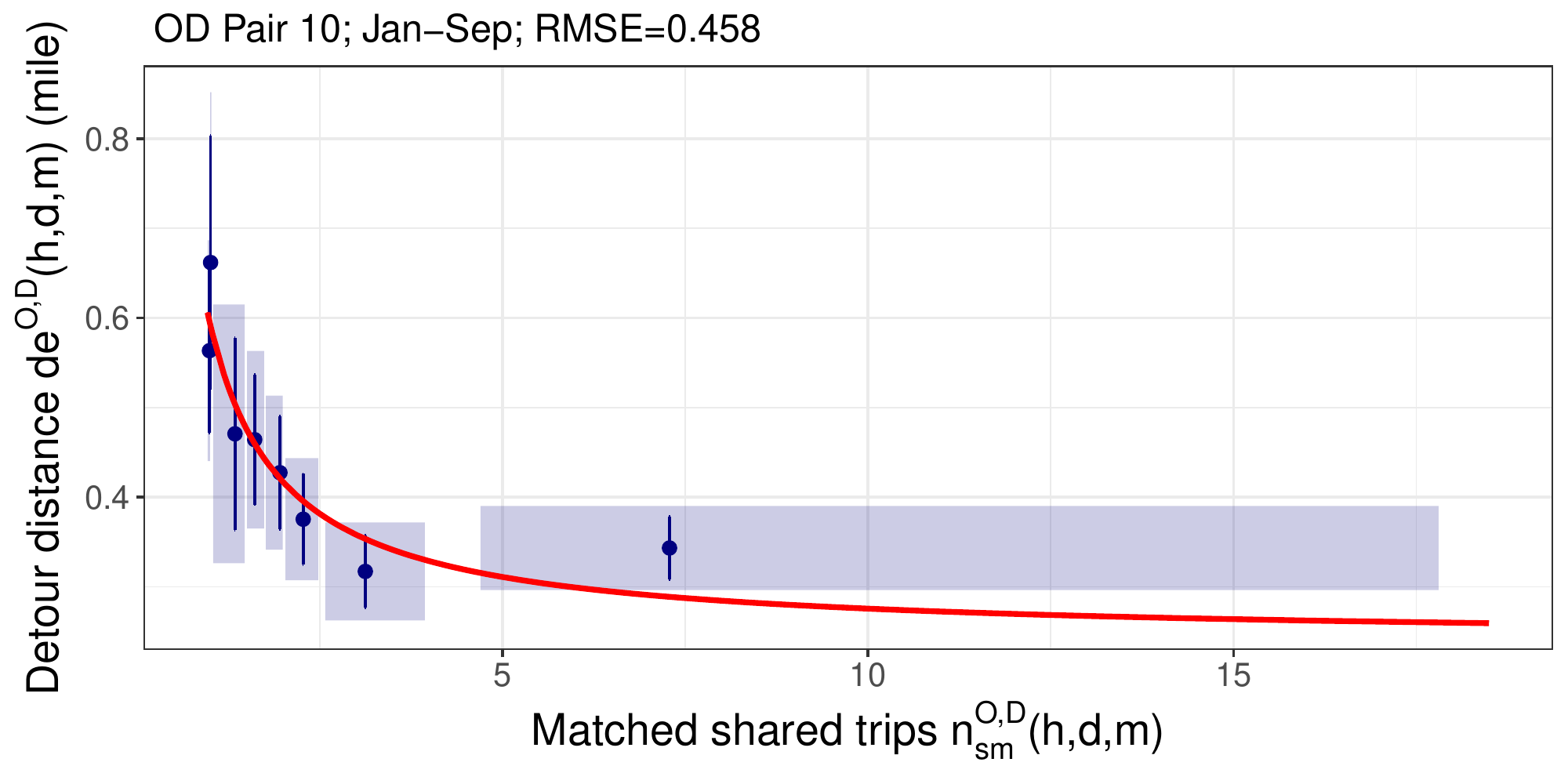}
    \end{subfigure}
\end{figure}%

\begin{figure}[H]\ContinuedFloat
    \centering
    \begin{subfigure}{0.45\textwidth}
        \centering
        \includegraphics[width=\textwidth]{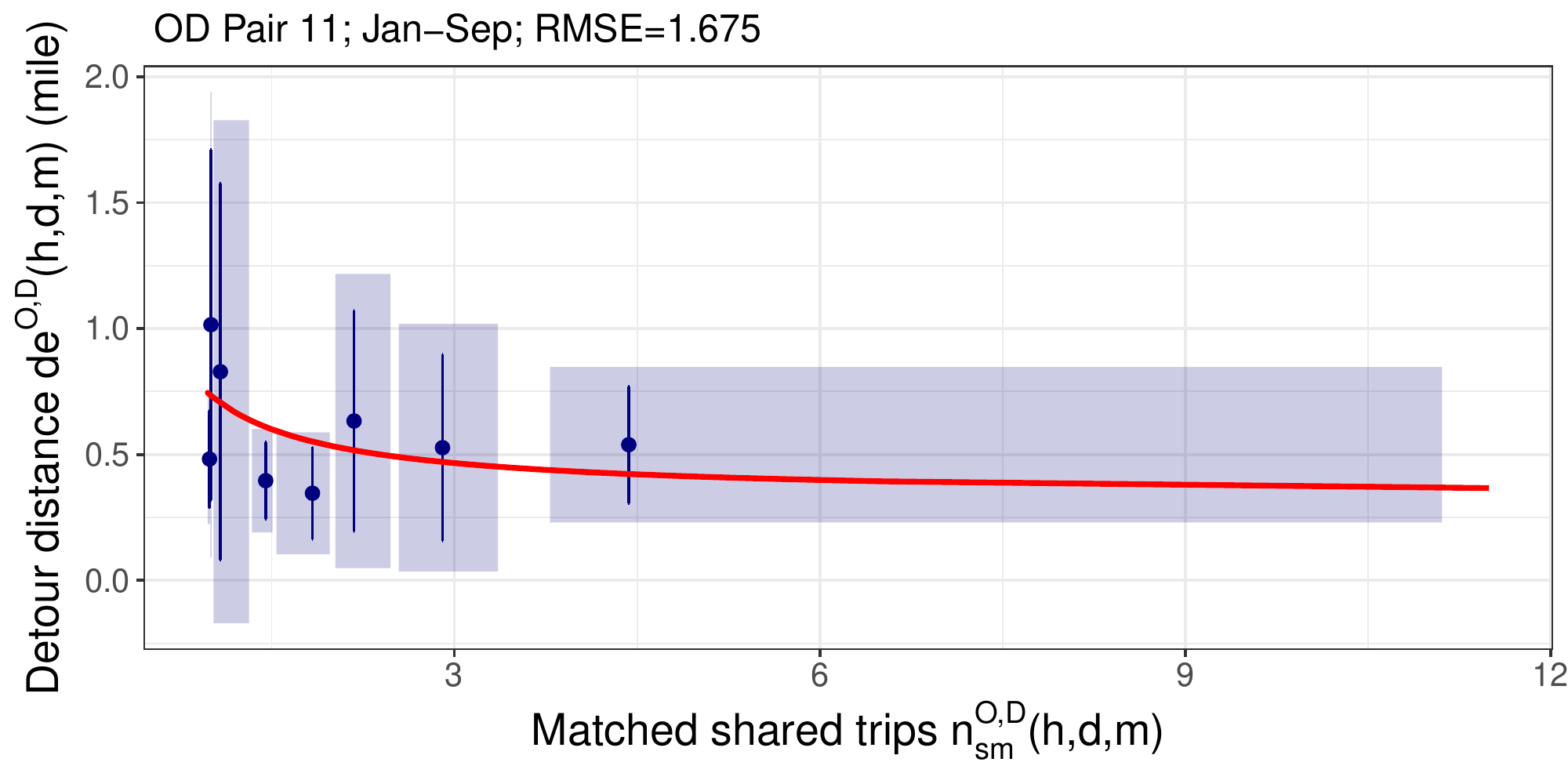}
    \end{subfigure}
    \begin{subfigure}{0.45\textwidth}
        \centering
        \includegraphics[width=\textwidth]{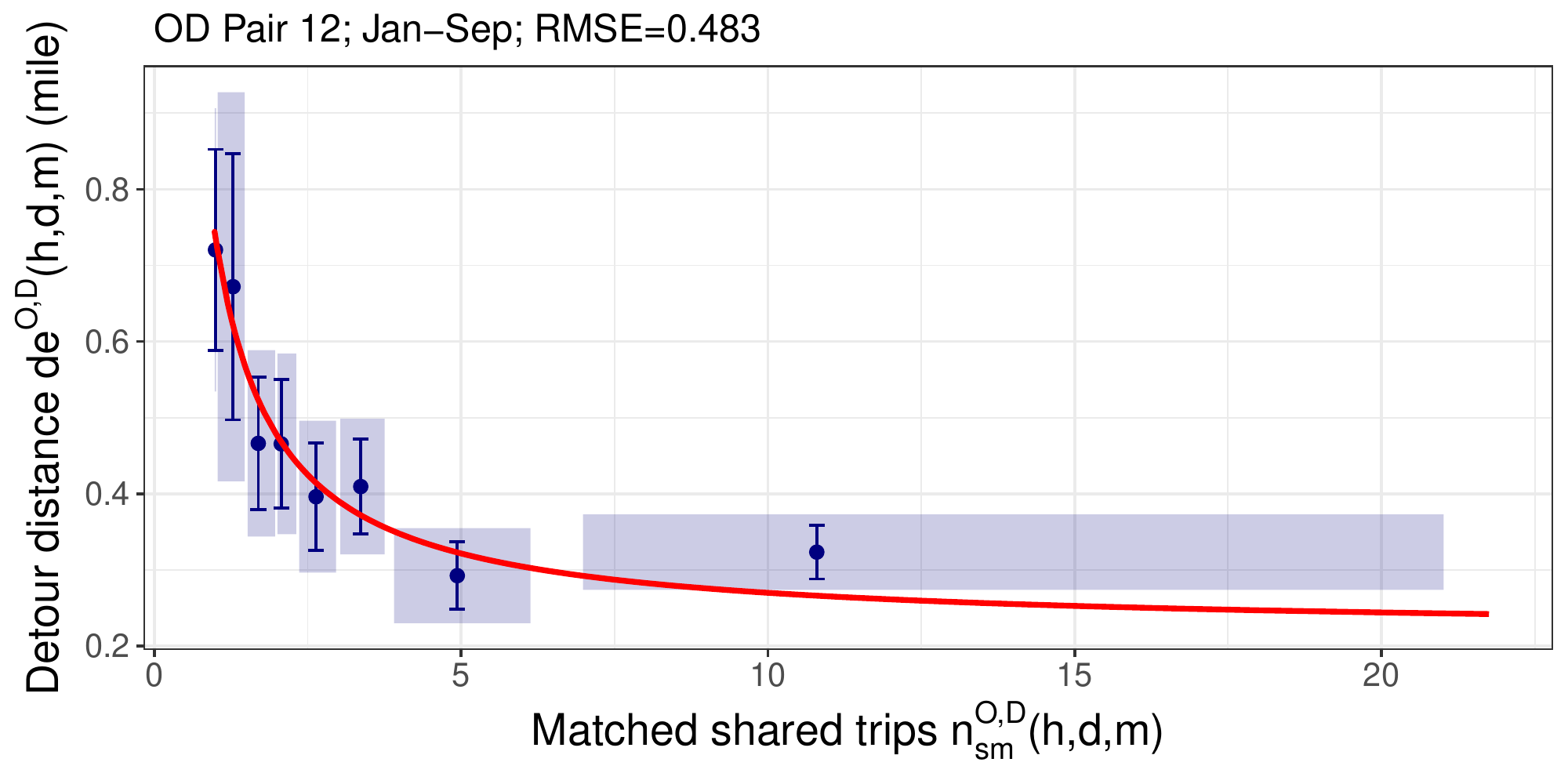}
    \end{subfigure}
    
    \begin{subfigure}{0.45\textwidth}
        \centering
        \includegraphics[width=\textwidth]{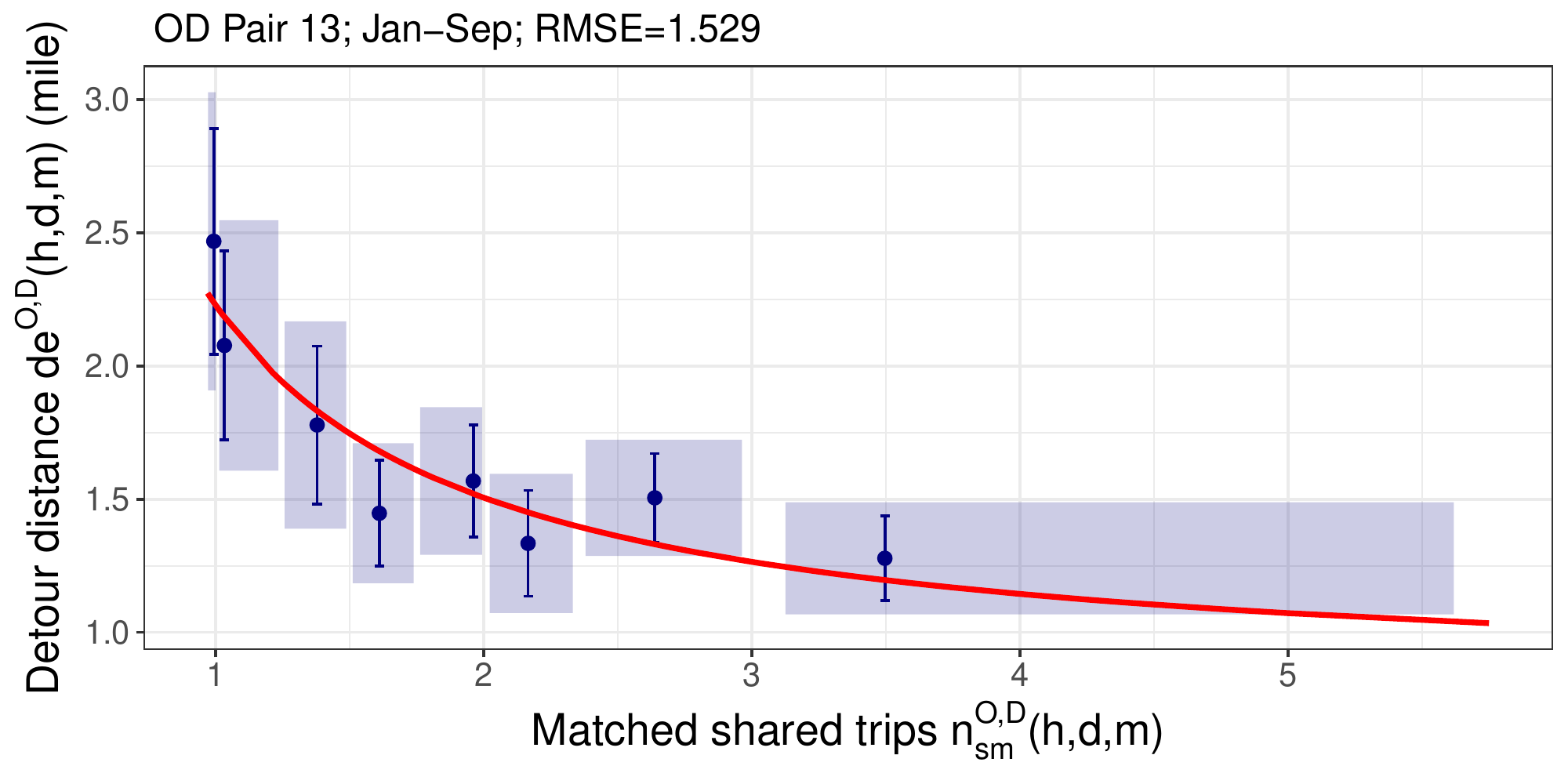}
    \end{subfigure}
    \begin{subfigure}{0.45\textwidth}
        \centering
        \includegraphics[width=\textwidth]{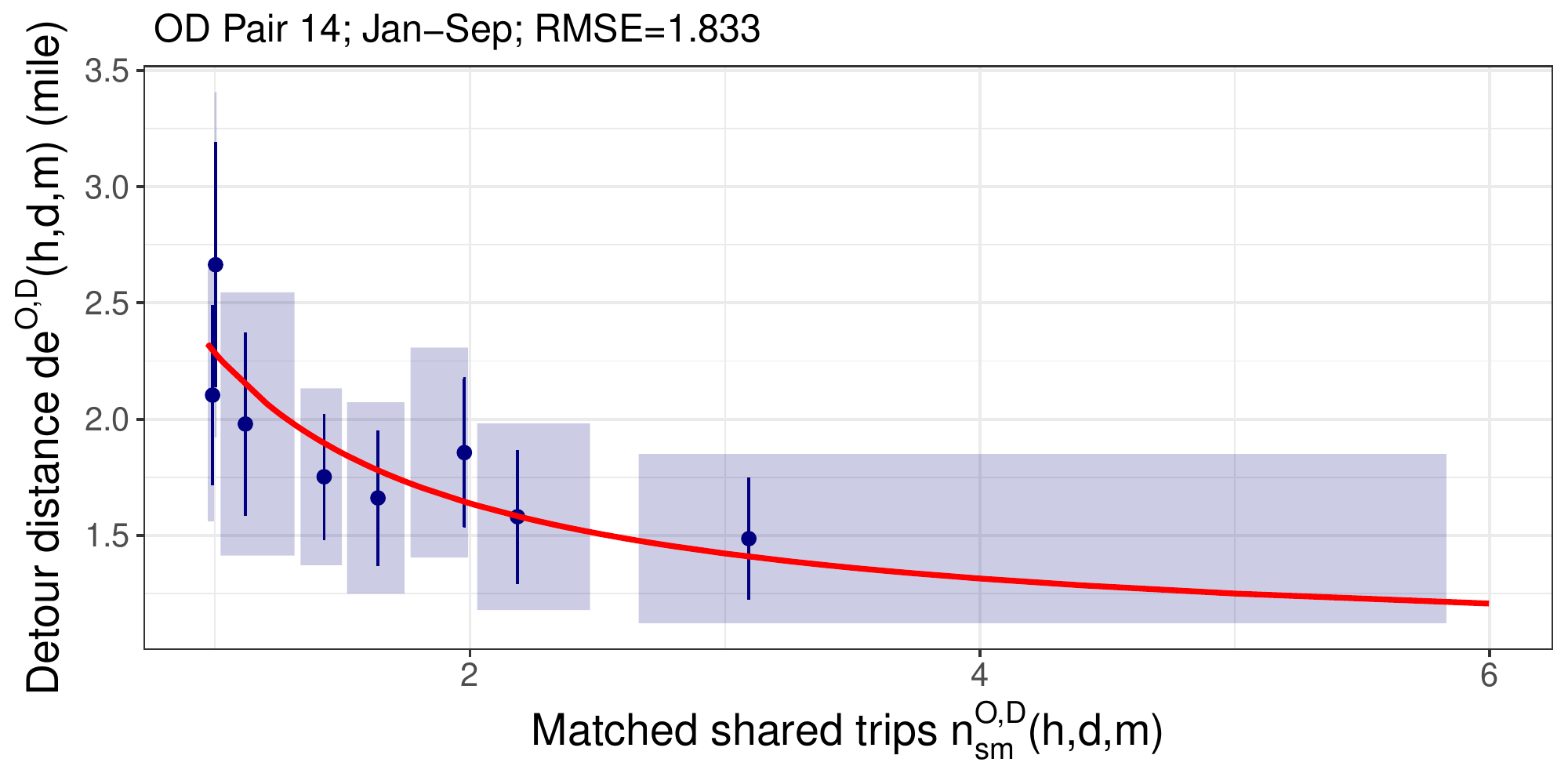}
    \end{subfigure}

    \begin{subfigure}{0.45\textwidth}
        \centering
        \includegraphics[width=\textwidth]{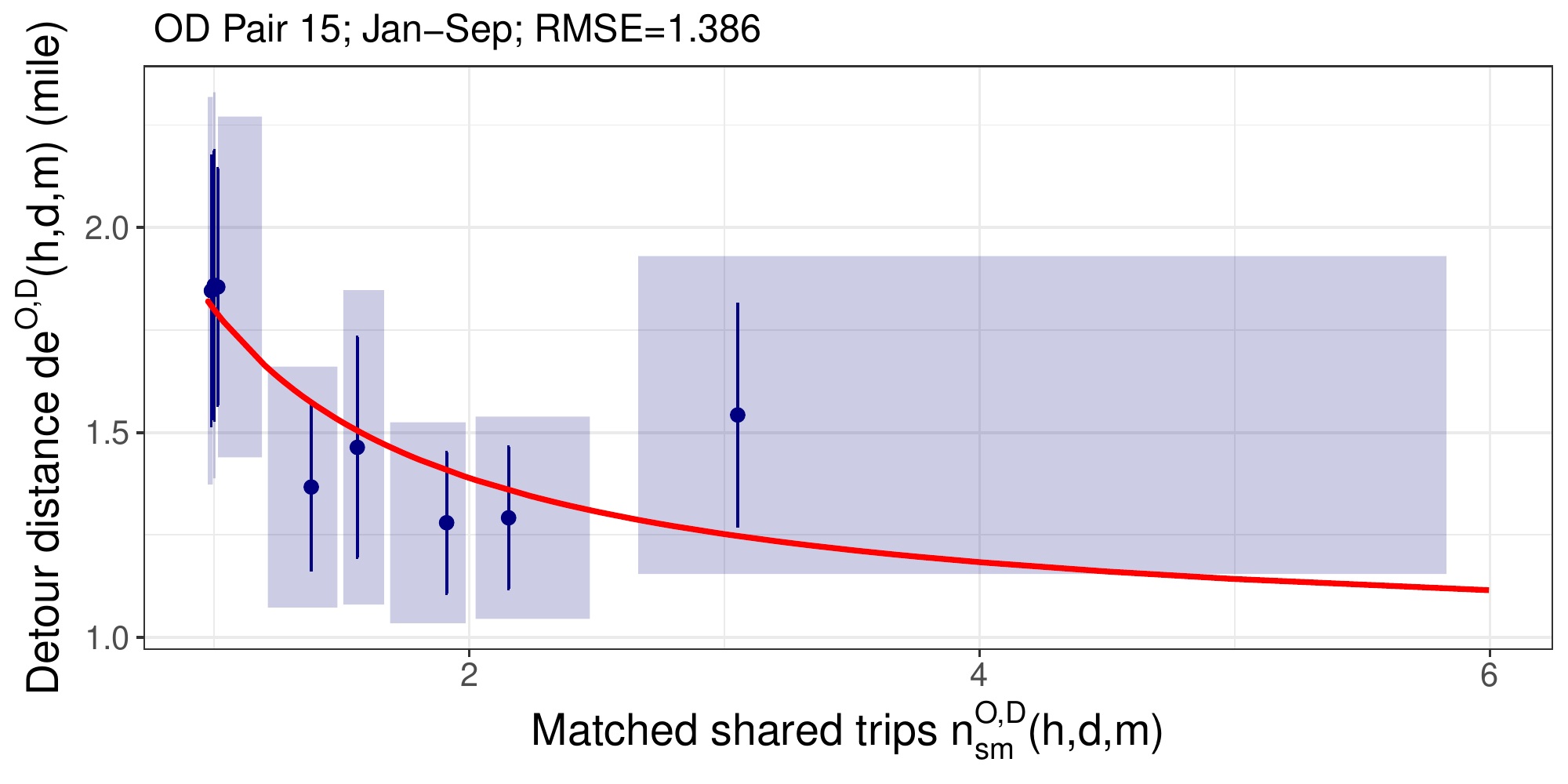}
    \end{subfigure}
    \begin{subfigure}{0.45\textwidth}
        \centering
        \includegraphics[width=\textwidth]{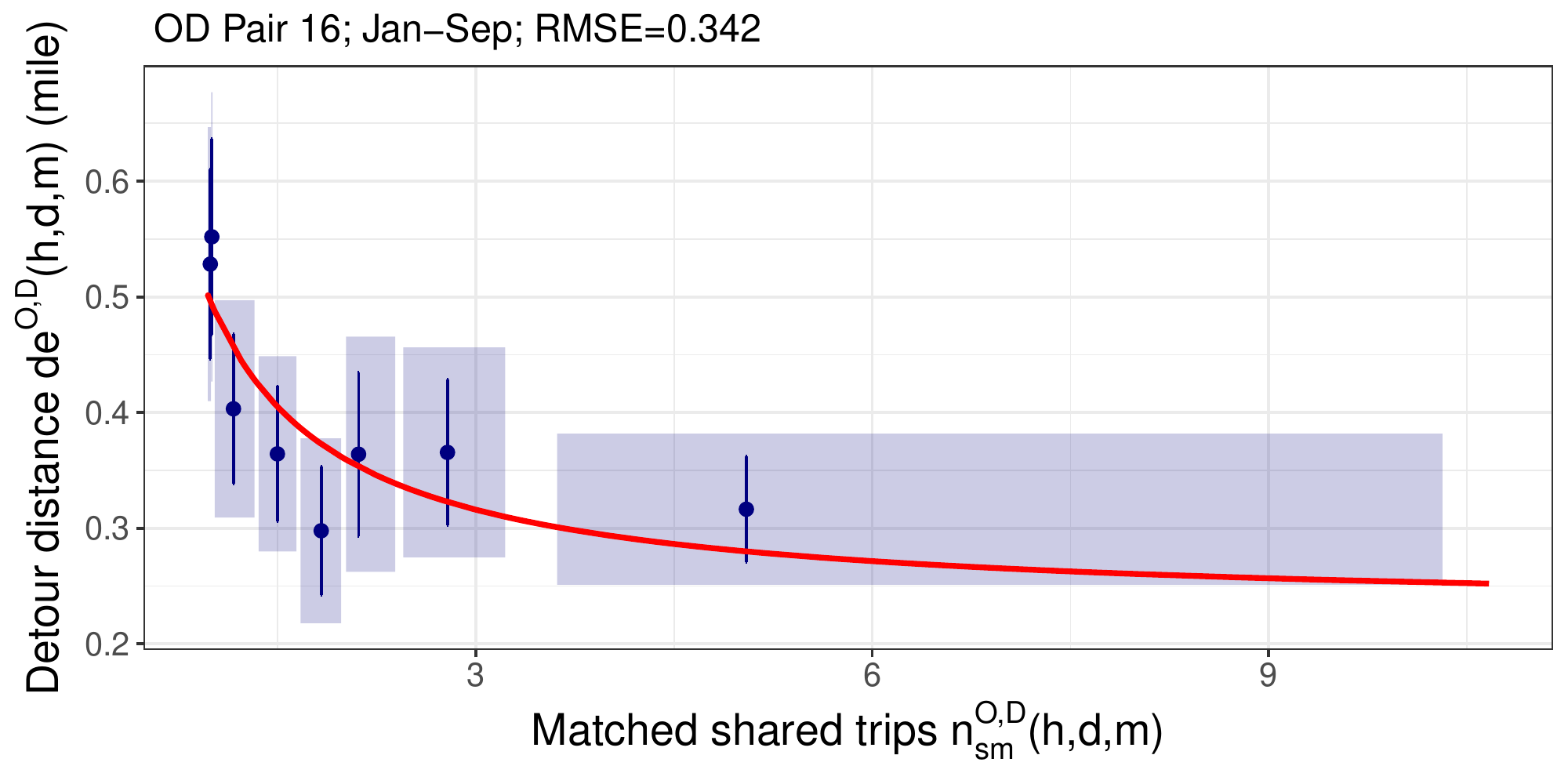}
    \end{subfigure}
    \caption{Detour distance for ten ODs.}
    \label{fig:detour_rest_10}
\end{figure}

The results of the regression for these 10 OD pairs are shown in Table \ref{tab:detour_disaggregate_7_16}. The elasticity of the detour distance for these 10 OD pairs varies between (-0.704, -0.452) and between (-0.321, -0.144) when the numbers of hourly matched shared trips between an OD pair are 1 and 5, respectively.

\begin{table}[!htbp] \centering 
  \caption{Regression results for detour distance (mile)} 
  \label{tab:detour_disaggregate_7_16} 
\begin{tabular}{@{\extracolsep{5pt}}lccccc} 
\\[-1.8ex]\hline 
\hline \\[-1.8ex] 
 & \multicolumn{5}{c}{\textit{Dependent variable:}} \\ 
\cline{2-6} 
\\[-1.8ex] & \multicolumn{5}{c}{Detour distance $de^{O,D}(h,d,m)$} \\ 
\\[-1.8ex] & (7) & (8) & (9) & (10) & (11)\\ 
\hline \\[-1.8ex] 
 $\frac{1}{n^{O,D}_{s_m}(h,d,m)}$ & 0.312$^{***}$ & 0.985$^{***}$ & 0.351$^{***}$ & 0.355$^{***}$ & 0.415 \\ 
  & (0.050) & (0.230) & (0.061) & (0.052) & (0.272) \\ 
  Constant & 0.293$^{***}$ & 0.927$^{***}$ & 0.166$^{***}$ & 0.240$^{***}$ & 0.324$^{*}$ \\ 
  & (0.032) & (0.181) & (0.044) & (0.035) & (0.195) \\ 
 \hline \\[-1.8ex] 
R$^{2}$ & 0.039 & 0.021 & 0.035 & 0.047 & 0.005 \\ 
Adjusted R$^{2}$ & 0.038 & 0.020 & 0.034 & 0.046 & 0.003 \\ 
\hline 
\hline 
\\[-1.8ex] & (12) & (13) & (14) & (15) & (16)\\ 
\hline \\[-1.8ex] 
 $\frac{1}{n^{O,D}_{s_m}(h,d,m)}$ & 0.519$^{***}$ & 1.434$^{***}$ & 1.298$^{***}$ & 0.820$^{***}$ & 0.266$^{***}$ \\ 
  & (0.061) & (0.194) & (0.257) & (0.190) & (0.043) \\ 
  Constant & 0.217$^{***}$ & 0.790$^{***}$ & 0.988$^{***}$ & 0.979$^{***}$ & 0.228$^{***}$ \\ 
  & (0.034) & (0.130) & (0.188) & (0.144) & (0.031) \\ 
 \hline \\[-1.8ex] 
R$^{2}$ & 0.088 & 0.052 & 0.029 & 0.021 & 0.045 \\ 
Adjusted R$^{2}$ & 0.087 & 0.051 & 0.028 & 0.020 & 0.044 \\ 
\hline 
\hline \\[-1.8ex] 
Significance levels & \multicolumn{5}{r}{$^{*}$p$<$0.1; $^{**}$p$<$0.05; $^{***}$p$<$0.01} \\ 
\end{tabular} 
\end{table} 

\clearpage
\bibliographystyle{elsarticle-harv}
\bibliography{biblio}
\end{document}